\documentclass[fleqn,usenatbib]{mnras}
\usepackage[T1]{fontenc}
\usepackage{ae,aecompl}

\usepackage{graphicx}	% Including figure files
\usepackage{amsmath}	% Advanced maths commands
\usepackage{amssymb}	% Extra maths symbols

\newcommand{\Msun}{M_{\odot}}

\newcommand{\tpb}{t_{\rm pb}}

\title[Systematic 3D MHD simulations of CCSNe - I.]{Three-dimensional Magneto-hydrodynamic Simulations of Core-collapse Supernovae: I. Hydrodynamic evolution and protoneutron star properties}

\author[K. Nakamura et al.]{
Ko Nakamura,$^{1,2}$\thanks{E-mail: nakamurako@fukuoka-u.ac.jp}
Tomoya Takiwaki$^{3}$
Jin Matsumoto$^{1,4}$
and Kei Kotake$^{1,2,5}$
\\
$^{1}$Department of applied physics, Fukuoka University, Nanakuma Jonan 8-19-1, Fukuoka 814-0180, Japan \\
$^{2}$Research Institute of Stellar Explosive Phenomena, Fukuoka University, Nanakuma Jonan 8-19-1, Fukuoka 814-0180, Japan \\
$^{3}$National Astronomical Observatory of Japan, Osawa 2-21-1, Mitaka, Tokyo 181-8588, Japan \\
$^{4}$Keio Institute of Pure and Applied Sciences, Keio University, Yokohama 223-8522, Japan\\
$^5$Institute for Theoretical Physics, University of Wroclaw, 50-204 Wroclaw, Poland
}

\date{Accepted 2024 November 13. Received 2024 November 13; in original form 2024 May 13}

\pubyear{2024}

\begin{document}
\label{firstpage}
\pagerange{\pageref{firstpage}--\pageref{lastpage}}
\maketitle

% Abstract of the paper
\begin{abstract}
We present results from three-dimensional, magnetohydrodynamic, core-collapse simulations of sixteen 
 progenitors following until 0.5\,s after bounce. 
We use non-rotating solar-metallicity progenitor models 
with zero-age main-sequence mass between 9 and 24 $\Msun$.
The examined progenitors cover a wide range of the compactness parameter 
including a peak around $23 \, \Msun$.
We find that 
neutrino-driven explosions occur for all models within 0.3\,s after bounce. 
We also find that the properties of the explosions and the central remnants are well correlated with the compactness.
Early shock evolution is sensitive to the mass accretion rate onto the central core, 
reflecting the density profile of the progenitor stars. 
The most powerful explosions with diagnostic explosion energy $E_{\rm dia} \sim 0.75 \times 10^{51}$ erg are obtained 
by 23 and 24 $\Msun$ models, which have the highest compactness among the examined models. 
These two models exhibit spiral SASI motions during 150--230 ms after bounce preceding a runaway shock expansion and leave a rapidly rotating neutron star with spin periods $\sim 50$ ms.
Our models predict the gravitational masses of the neutron star ranging between $1.22 \Msun$ and $1.67 \Msun$ and their spin periods 0.04\,--\,4\,s. 
The number distribution of these values roughly matches observation. 
On the other hand, our models predict small hydrodynamic kick velocity (15 -- 260\,${\rm km \, s}^{-1}$), although they are still growing at the end of our simulations. 
Further systematic studies, including rotation and binary effects, 
as well as long-term simulations up to several seconds, 
will enable us to explore the origin of various core-collapse supernova explosions.
\end{abstract}

% Select between one and six entries from the list of approved keywords.
% Don't make up new ones.

%%%%%%%%%%%%%%%%%%%%%%%%%%%%%%%%%%%%%%%%%%%%%%%%%%

%%%%%%%%%%%%%%%%% BODY OF PAPER %%%%%%%%%%%%%%%%%%

\section{Introduction} \label{sec:intro}
Accounting for fundamental properties of core-collapse supernova (CCSN) observations, such as the explosion energy, synthesized nickel mass, and mass of the central remnant, provides critical benchmarks for verifying the validity of the theoretical CCSN modeling. 
By comparing these observational evidence with numerical outcomes of CCSN simulations, it is expected that one can place tight constraints on massive stellar evolutionary path, in principle, from the zero-age main sequence stage, to the final explosion or collapse 
of the massive stars.
Furthermore, predicting CCSN multi-messenger signals, such as neutrinos, gravitational waves, and electromagnetic signals, is crucial to maximize the scientific return from a once-in-a-lifetime event -- Galactic supernova.

To explore the general properties of CCSN explosions, many efforts have been devoted to systematic CCSN studies covering a wide range of the progenitor's mass and metallicity. 
One-dimensional (1D) simulations assuming spherical symmetry to reduce computational costs have been suitable for this purpose. 
By performing general-relativistic (GR) simulations for over 100 presupernova models using a leakage scheme, 
\citet{Oconnor11a} 
were the first to 
point out that the postbounce dynamics and the 
progenitor-remnant connections are predictable basically
by a single parameter, compactness $\xi$ of the stellar core at bounce 
(see also \cite{oconnor13}).
\cite{ebinger18} also shows the explodability of progenitors with nucleosynthesis prediction using a neutrino radiation hydrodynamic scheme with an enhancement of the heating rate of heavy lepton neutrino.
Along this line, \citet{Ugliano12} 
performed 1D hydrodynamic simulations for 101 progenitors of \citet{WHW02}.
By replacing the proto-neutron star (PNS) interior with an inner 
boundary condition, they followed an unprecedentedly 
long-term evolution over hours to days after bounce in spherical symmetry. 
Their results also lent support to the finding by \citet{Oconnor11a} 
that the 
compactness parameter is a good measure for diagnosing the progenitor-explosion and the progenitor-remnant correlation.

Instead of the one parameter $\xi$, \citet{ertl16} proposed two parameters, $M_4$ and $M_4\mu_4$, and they insisted that the two-parameter criterion can separate successful explosions from failures with very high reliability. 
These 1D studies have been hugely extended by \citet{sukhbold16,Sukhbold18}, who calculated thousands of pre-SN evolution and concluded that
the explodability -- a star explodes or not --- may reflect the small, almost random differences in its late evolution more than its initial mass. 

To construct 1D CCSN models, some artificial boost to drive explosion is inevitable 
since they can not take account of multi-dimensional (multi-D) effects such as 
neutrino-driven convection 
\citep[e.g.,][]{burrows95,Janka96}
and the standing-accretion-shock-instability 
\citep[SASI;][]{Blondin03,Foglizzo06,Foglizzo07,Ohnishi06,Blondin07_nat,Iwakami08,Iwakami09,couch13},
which play essential roles in neutrino-driven explosions. 
1D models denoted above employ artificially enhanced neutrino heating rate or analytic formula for neutrino luminosity as a function of time, calibrated by SN 1987A.

\citet{nakamura15} included these multi-D phenomena in a self-consistent manner by performing two-dimensional (2D) core-collapse simulations. 
They conducted neutrino-radiation hydrodynamics calculations for about 400 progenitor models covering a wide range of mass (10.8 -- $75\,\Msun$) and metallicity (zero to solar) and confirmed that CCSN explosion properties such as explosion energy, neutrino luminosity, remnant mass, and mass of synthesized nickel in the ejecta are well correlated to the compactness parameter. 
Furthermore, they extended some of their models to longer timescales and found a high neutron star kick velocity driven by aspherical mass ejection in a high compactness progenitor \citep{nakamura19}.
\cite{summa16} also performed 2D study and the importance of Si/O interface is discussed. \cite{vartanyan23}
used an elaborated neutrino transport scheme and systematically showed the dependence of the neutrino emission property on the progenitor structure. 
The effects of multi-dimensional turbulence have been phenomenologically treated in several 1D simulations \citep{mullerb19,couch20,boccioli21,boccioli23,sasaki24}. 
That significantly reduces the computational costs of multi-dimensional simulations and eases the systematic study.

Although 2D simulations can take account of the effects of the hydrodynamic instabilities, 2D models tend to overestimate the explodability since a powerful sloshing mode of SASI emerges along the symmetry axis. 
Moreover, there are some phenomena unique to three-dimensional (3D) models, such as a spiral mode of SASI and Lepton-number Emission Self-sustained Asymmetry \citep[LESA;][]{tamborra14b}, 
which might affect CCSN dynamics and multi-messenger signals.

The number of 3D CCSN models has been growing, 
aided by the development of high-performance computers and numerical schemes. 
Along the lineage of the systematic CCSN studies, 
\citet{burrows20} conducted 3D core-collapse simulations for fourteen solar-metallicity progenitors from \citet{sukhbold16} 
and an additional subset with different physical treatments (see also \citealt{vartanyan19b}).
Interestingly, some of their models in the middle of their available progenitor mass range, specifically 13, 14, and $15\,\Msun$ models, do not explode in their simulation time. 
Since the compactness parameters of these models are smaller than the exploding higher mass models, 
they concluded that 
the compactness measure is not a metric for explodability, 
while high-$\xi$ exploding models present high growth rates of the explosion energy. 
They argues that density drop rate at Si/O interface shows better correlation with the explodabiility \citep{wang-tanshu22,tsang22}.

To understand the connection between the progenitor structure and explosion properties, independent investigations based on 3D CCSN modelling are necessary. 
We have performed, for the first time, systematic 3D magnetohydrodynamic (MHD) core-collapse simulations up to 0.5 s after bounce. 
Our simulations cover a mass range of the progenitor stars from $9\,\Msun$ to $24\,\Msun$, including a peak of the compactness parameter around $23\,\Msun$. 
We note that the initial magnetic strength assumed in our model ($B_0 = 10^{10}$ G at the centre) is weak and does not play any significant role in dynamic evolution \citep[see][for the $B_0$ dependence of the shock evolution]{matsumoto20,matsumoto22}.

Section~\ref{sec:setup} describes the numerical setup, 
including an explanation of our numerical scheme and
structure of the progenitor models. 
Results start from Section~\ref{sec:hydro}. 
In this paper, we report the results of our 3D MHD core collapse simulations focusing on their hydrodynamic evolution (Section~\ref{sec:hydro}) and the resultant remnant properties (Section~\ref{sec:pns}). 
In Section~\ref{sec:mag} we briefly discuss the impact of the weak magnetic field on the hydrodynamic evolution. 
We summarize our results and discuss their implications in Section~\ref{sec:summary}.

\section{Numerical setups} \label{sec:setup}

The \texttt{3DnSNe} code we employ in this study is a multi-dimensional, three-neutrino-flavour radiation hydrodynamics code constructed to study core-collapse supernovae. 
We have updated this code to deal with magnetohydrodynamics (MHD) employing a divergence cleaning method with careful treatments of finite volume and area reconstructions \citep{matsumoto20}.  
The initial magnetic field is given by a vector potential in the $\phi$-direction of the form
\begin{equation}
A_\phi = \frac{B_0}{2} \frac{r_0^3}{r^3 + r_0^3} r \sin \theta,
\end{equation}
where $r_0 = 1000$ km characterizes the topology of the field. 
The magnetic field is uniform when the radius $r$ is smaller than $r_0$, 
while it is like a dipole field when $r$ is larger than $r_0$. 
$B_0$ determines the strength of the magnetic field inside the core ($r < r_0$). 
In this study, we fix the strength of the initial magnetic field to be $B_0 = 10^{10}$ G. 

\begin{figure}
    \includegraphics[width=0.95\linewidth]{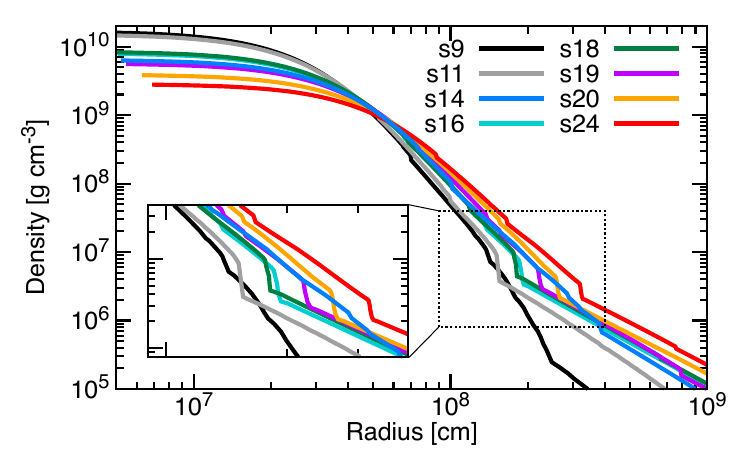} 
    \caption{
    Density distribution of the progenitor models as a function of radius. The density jumps located at 1500--3000 km correspond to the interface of the Si/O layers. 
    To improve readability, we show eight progenitors from the sixteen progenitors examined in this study and highlight the profile around the density drops in the inset. Refer to Figure \ref{fig:prog-den-all} for the structure of the remaining models not shown here.
    }
    \label{fig:prog-den}
\end{figure}

We solve the neutrino transport by the isotropic diffusion source approximation (IDSA) scheme 
for electron, anti-electron, and heavy lepton neutrinos \citep{idsa,takiwaki16} 
with discretized neutrino spectrum described by 20 energy bins for $0 < \epsilon_\nu \leq 300$\,MeV. 
The neutrino reaction rate is almost the same as ``set-all'' of \citet{kotake18}, except for the removal of the reactions of electron–neutrino pair annihilation into $\mu$/$\tau$ neutrinos and $\mu$/$\tau$–neutrino scattering on electron (anti)neutrinos to save computational resources. Refer to \citet{matsumoto22} for details on this treatment.
\citet{takiwaki18} have implemented the gravity potential taking account of the effective General Relativistic effect 
\citep[case A in][]{marek06}.

We simulate the core collapse and bounce in 2D geometry to reduce the computational costs, 
and then map it to the 3D coordinates at 10\,ms after bounce. 
We employ a spherical grid in $r$ and $\theta$ ($r$, $\theta$, and $\phi$) of resolution $600 \times 128$ ($600 \times 64 \times 128$) for our 2D (3D) simulations. 
The spatial range of the computational domain is within $r <$10,000\,km in radius and divided into 600 non-uniform radial zones. Our spatial grid has the finest mesh spacing $\Delta r_{\rm min} = 250$\,m at the centre, and $\Delta r/r$ is better than 1.0\,\% at $r > 100$\,km. 
The polar angle grid ($0 \leq \theta \leq \pi$) is given by $\Delta(\cos \theta)$ = const. to avoid restrictive CFL timestep limitations, whereas the azimuthal angle ($0 \leq \phi \leq 2\pi$) is uniformly divided.
A special treatment is imposed in the innermost 10 km where 
non-radial motions are suppressed 
to avoid excessive time-step limitations. 
Seed perturbations for aspherical instabilities are imposed by hand at the time of 3D mapping by introducing random perturbations of $0.1\%$ in density on the whole computational grid except for the unshocked core.
Regarding the equation of state (EOS), we use that of \citet{lattimer91} 
with a nuclear incompressibility of $K = 220$\,MeV.
At low densities, 
we employ an EOS accounting for photons, electrons, positrons, 
and ideal gas contribution. 

The progenitor models we use are solar-metallicity single stars from \citet{sukhbold16}. 
We investigate sixteen progenitor models in increments of one solar mass from $9\,\Msun$ to $24\,\Msun$.
We designate models with the solar metallicity as s models and label the models with their initial ZAMS mass. For example, an s model of $9\,\Msun$ is referred to as s9. 
Figure~\ref{fig:prog-den} describes the density profile of the progenitor models. 
Most progenitors present conspicuous density drops at 1500--3000\,km in radius, corresponding to a discontinuity of chemical compositions at the Si/O interface.
On the other hand, the model s14 (blue line in Figure~\ref{fig:prog-den}) has the Si/O interface at $\sim \, 3000$ km, but the density drop there is negligible.
These features are critical to shock evolution, as discussed later.
Note that in the following, we will show not all but some selected models as representatives (for example, eight of sixteen models for Figure~\ref{fig:prog-den}) 
so that the differences between the models are distinguishable. 
The plots including the models not shown in the main text appear in Appendix~\ref{sec:restandfullsetmodels}.

\begin{figure}
    \begin{center}
    \begin{tabular}{c}
    \includegraphics[width=0.95\linewidth]{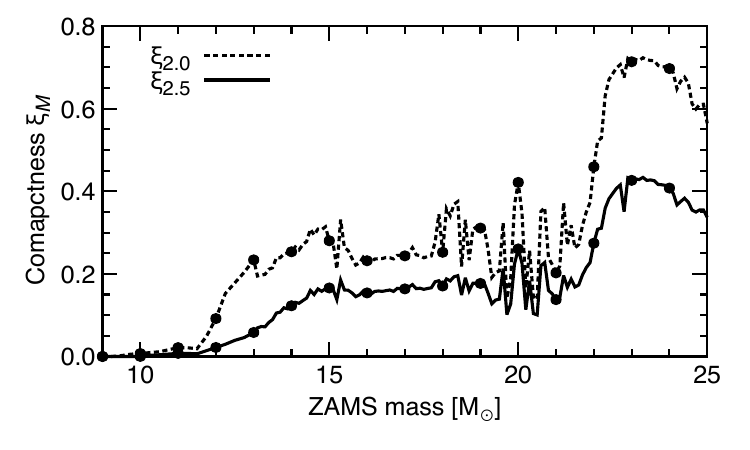} \\
    \includegraphics[width=0.95\linewidth]{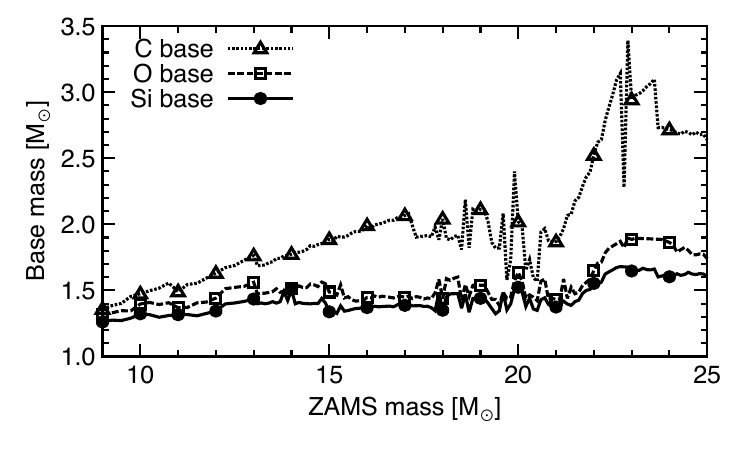} \\   \includegraphics[width=0.95\linewidth]{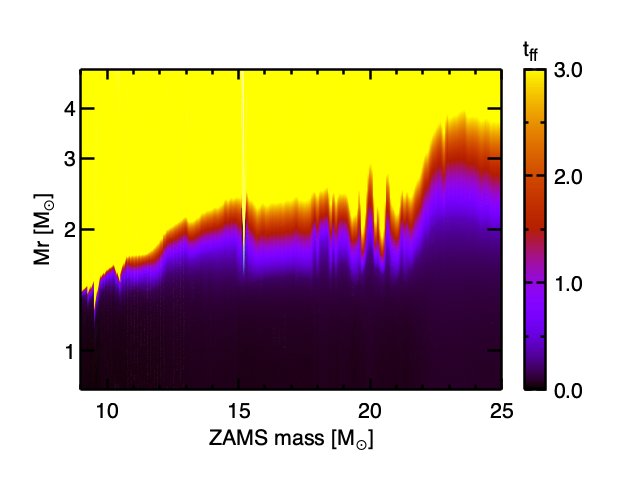}
    \end{tabular}
    \end{center}
    \caption{The structure of the progenitor models from \citet{sukhbold16}.
    Top panel: the compactness parameters $\xi_{2.0}$ and $\xi_{2.5}$ defined in Eq.~\eqref{eq:xi} for zero-age main-sequence (ZAMS) mass between $9\,\Msun$ and $25\,\Msun$.
    The examined models in this article, indicated by filled circles, cover a wide range of the compactness 
    from the smallest $\xi_{2.5}$ ($3.8 \times 10^{-5}$ of s9) 
    to almost the highest value ($0.43$ of s23) in this mass domain.
    Middle panel: distribution of mass coordinates at the base of the silicon layer (solid line), the oxygen layer (dashed line), and the carbon-rich layer (dotted line). 
    The base of the silicon, oxygen, and carbon layers are defined by the conditions of $X(^{28}{\rm Si}) = 10^{-2}$, $X(^{16}{\rm O}) = 10^{-3}$, and $X(^{12}{\rm C}) = 10^{-3}$, respectively. 
    The symbols on each line highlight the sixteen models examined in this paper. 
    Bottom panel: distribution of free-fall timescale $t_{\rm ff}(M)$ estimated by Eq.~\eqref{eq:tff} as a function of the mass coordinate. 
    }
    \label{fig:prog}
\end{figure}

The progenitor's compactness $\xi_M$ is a good diagnostic parameter for the explosion properties.
It is defined in \citet{Oconnor11a} 
as a function of an enclosed mass $M$, 
\begin{equation}
\xi_M = \frac{M \, [\Msun]}{R(M) \, [1000 \, {\rm km}]}.
\end{equation}\label{eq:xi}
The top panel of Figure~\ref{fig:prog} shows the compactness $\xi_M$ 
of the Sukhbold's progenitor models 
for $M=2.0 \Msun$ and $2.5 \Msun$ as a function of the progenitors' ZAMS mass. 
The compactness $\xi_M$ is roughly proportional to the ZAMS mass up to $\sim 15 \Msun$, almost constant in the range of $15 \Msun \lesssim M_{\rm ZAMS} \lesssim 18 \Msun$, very fluctuated around $M_{\rm ZAMS} \sim 20 \Msun$, and peaked at $22 \Msun \lesssim M_{\rm ZAMS} \lesssim 25 \Msun$. 
This non-monotonic profile of the compactness $\xi_M$ is mainly caused by the ignition and duration time of C-shell burning. See \cite{Takahashi2023} for the detailed discussions. 
The models examined in this paper are denoted by symbols in Figure \ref{fig:prog}. It can be seen that the selected models cover a wide range of the compactness parameter including the peak at $\sim 23\,\Msun$.

Following \citet{Takahashi2023}, we estimate the mass coordinates of bases of chemically defined layers (Si, O, and C) to characterize the chemical distribution of the progenitor models.
The base of each layer is defined at the innermost mass coordinate where the mass fraction of the element firstly exceeds a certain limit 
(e.g., $X(^{28}{\rm Si}) = 10^{-2}$).
They roughly correspond to the surfaces of Fe, Si, and O cores. 
It can be seen from the middle panel of Figure \ref{fig:prog} that the progenitor models with ZAMS mass around $\sim 23\,\Msun$ have massive Fe cores surrounded by dense Si-rich and O-rich layers. 

The compactness parameters $\xi_M$ in Figure \ref{fig:prog} are estimated from the pre-collapse progenitor data. 
Note that $\xi_M$ depends on the estimated time since the core structure changes drastically during the gravitational collapse \citep{oconnor13,Takahashi2023}. 
We estimate a free-fall time scale $t_{\rm ff}(M)$ of each pre-collapse progenitor star defined by 
\begin{equation}\label{eq:tff}
    t_{\rm ff}(M) = \frac{\pi}{2\sqrt{2}}\sqrt{\frac{R(M)^3}{GM}},
\end{equation}
at the mass coordinate $M$ (the bottom panel of Figure~\ref{fig:prog}). We confirm that $t_{\rm ff}(M)$ for $M \gtrsim 2.0 \Msun$ is equivalent to or longer than our computation time (0.65--0.75\,s) 
for most of the progenitor models examined in this paper. 
For high-$\xi$ models $t_{\rm ff}(M<2\,\Msun)$ is small (e.g., $t_{\rm ff}(M=1.75\,\Msun)=0.21$\,s for the s24 model), which means that their explosion properties could be largely affected by the matter initially located far from the center. Since the compactness parameter exclusively reflects structures within $M$, careful consideration should be given to the selection of $M$ when investigating the correlation between the explosion properties and $\xi_M$.

Along with these setups, we simulate core collapse, bounce, and subsequent shock evolution 
under the influence of neutrino heating and magnetism. 
Our simulations are self-consistent 
in the sense that we do not employ any artificial control to drive explosions. 

For reference, we conduct additional simulation without magnetism ($B_0 = 0$) for the $23\,\Msun$ model to assess the impact of the (weak) magnetic field on hydrodynamic evolution (Section~\ref{sec:mag}).

\section{Hydrodynamic evolution}\label{sec:hydro}

In this section we investigate the dynamical evolution of our CCSN models. Section \ref{sec:overview} overviews the morphology of the shock wave and the matter behind the shock at the final timestep of our simulations. 
In the subsequent sections the time evolution of the shock wave (Section~\ref{sec:shock}) and the explosion energy (Section~\ref{sec:energy}) are discussed.

\subsection{Overview}\label{sec:overview}

\begin{figure*}
    \begin{center}
    \begin{tabular}{ccc}
    \includegraphics[width=0.33\linewidth]{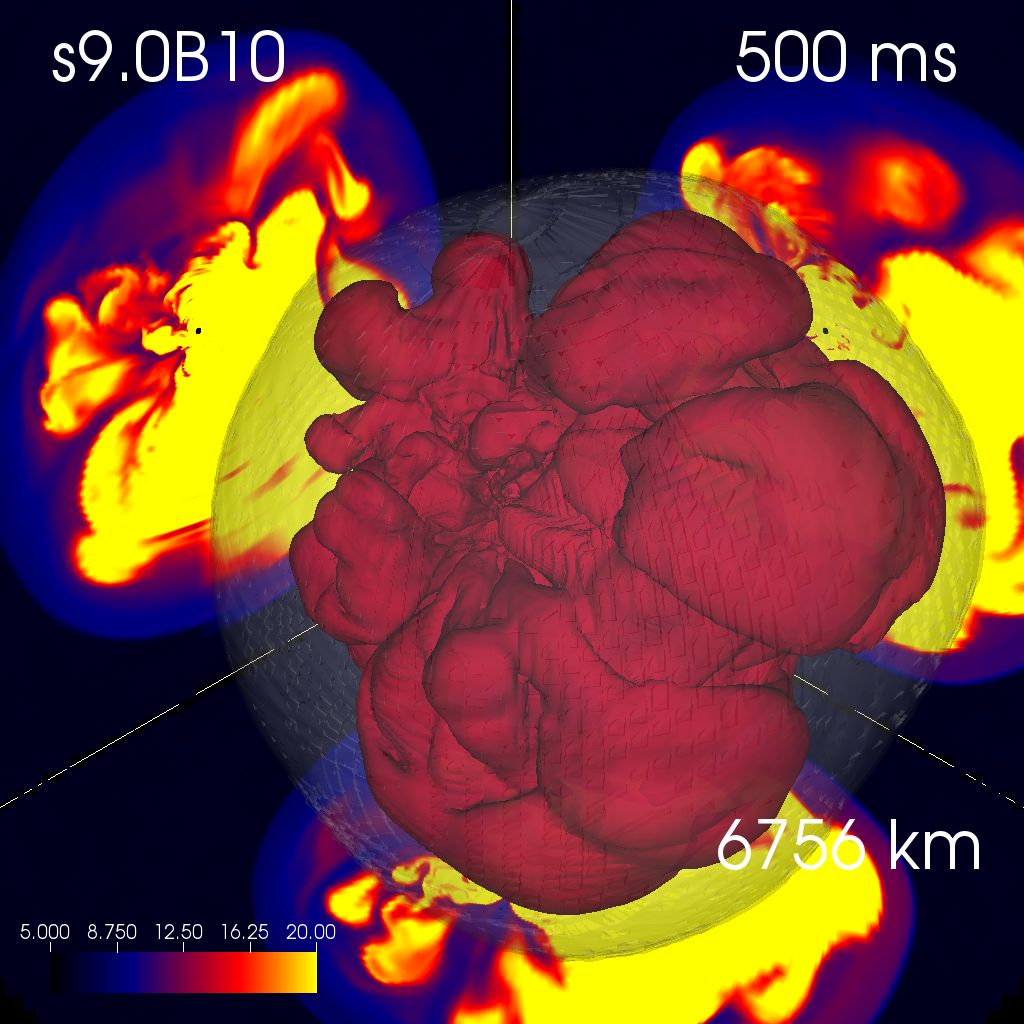} &
    \includegraphics[width=0.33\linewidth]{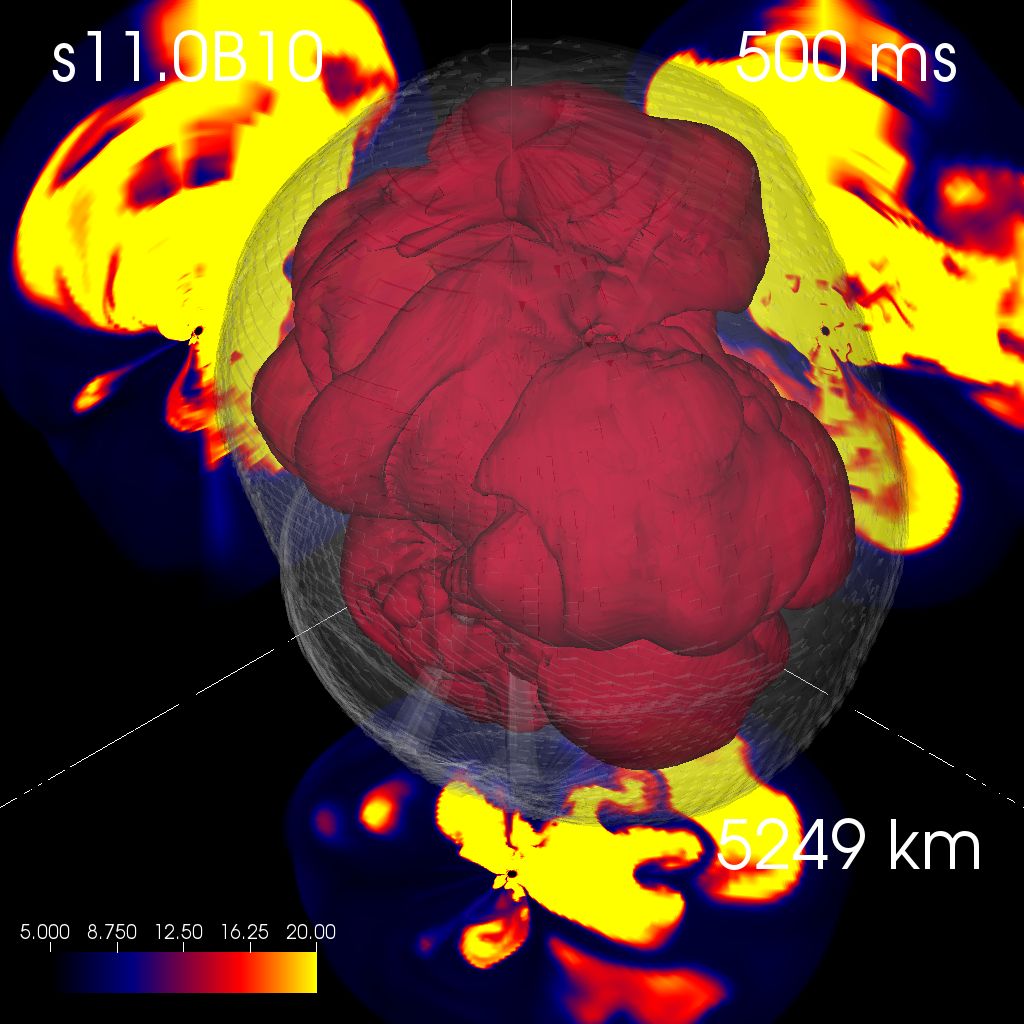} &
    \includegraphics[width=0.33\linewidth]{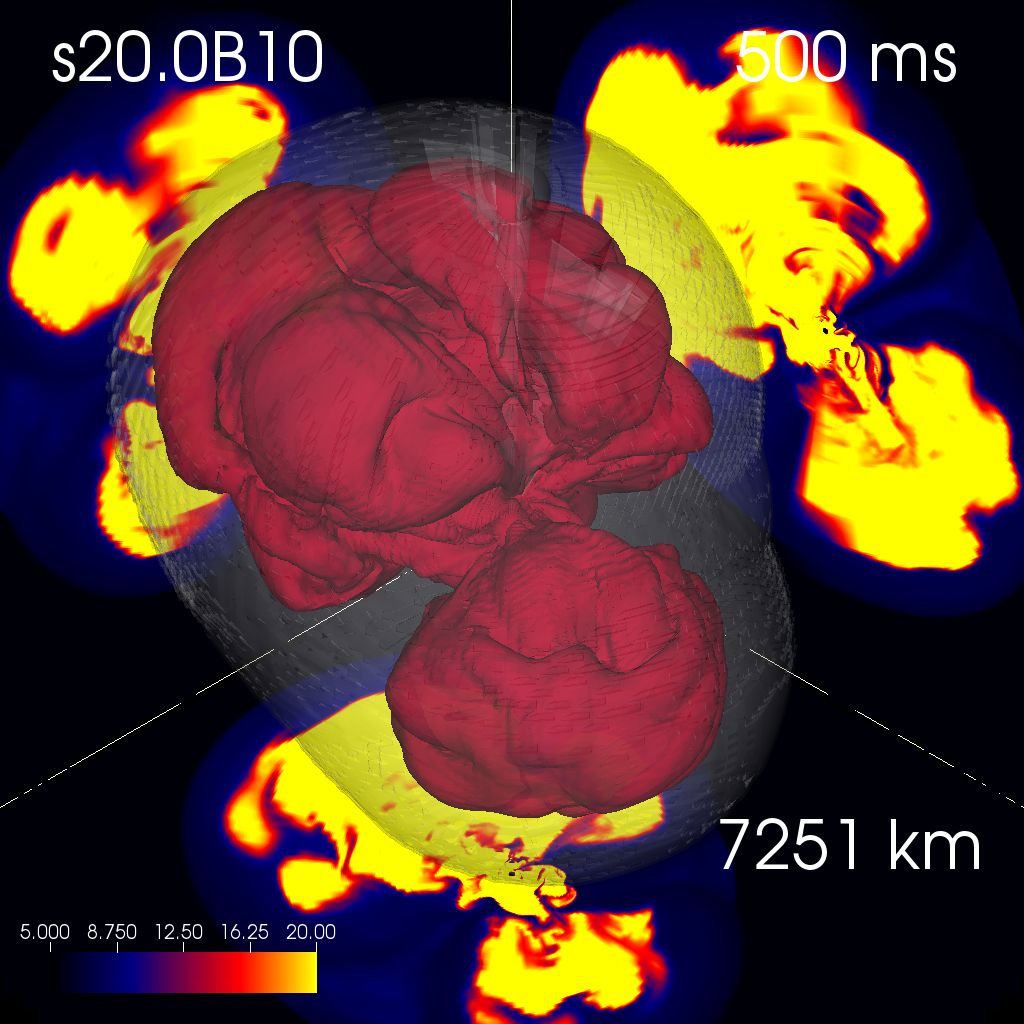} \\

    \includegraphics[width=0.33\linewidth]{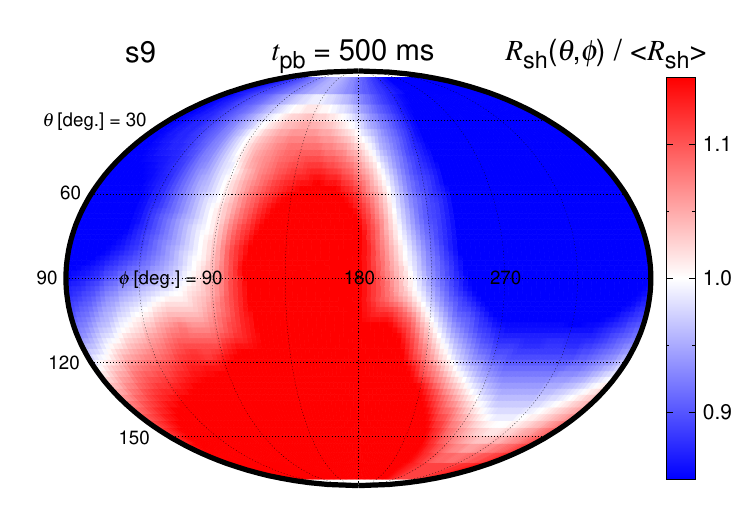} &
    \includegraphics[width=0.33\linewidth]{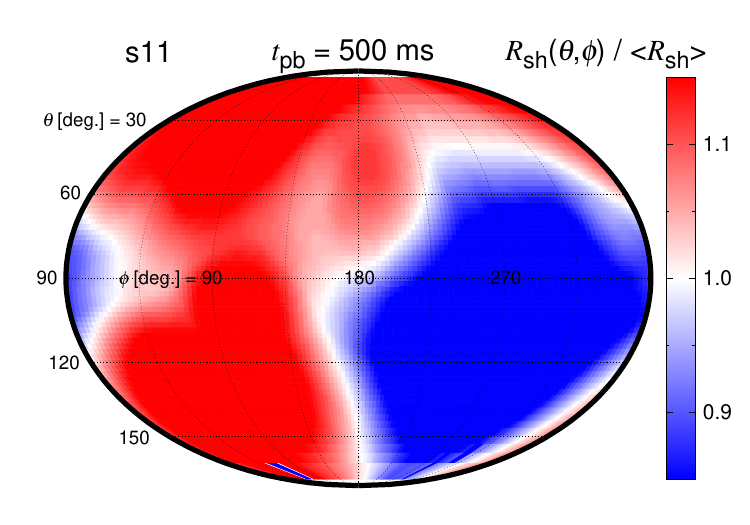} &
    \includegraphics[width=0.33\linewidth]{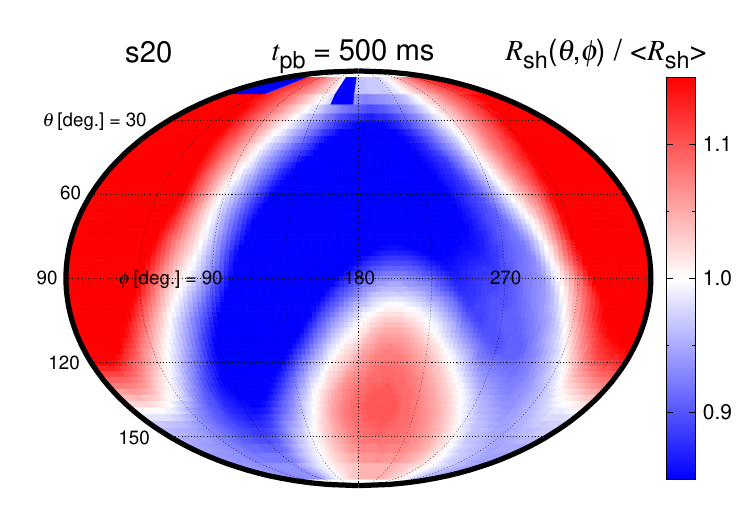} \\
    
    \end{tabular}
    \end{center}
    \caption{Top row: snapshots of the isosurface of entropy at 500\,ms after bounce for the models s9 (left panel), s11 (middle panel), and s20 (right panel). The scale of the visualized box is shown in each panel. The contours on the cross-sections in the $x=0$, $y=0$, and $z=0$ planes are projected on the walls at the back right, back left, and bottom of each panel, respectively.
    Bottom row: Mollweide-like maps of the angle dependent shock radius $R_{\rm sh}(\theta,\phi)$ to the $4\pi$ average.}
    \label{fig:snap}
\end{figure*}

Figure~\ref{fig:snap} shows the snapshots of the isosurface of entropy at 500 ms after bounce for the models s9, s11, and s20.
Our simulations start from spherical geometries except the dipole-like magnetic field. 
After the core bounce, hydrodynamic non-spherical instabilities develop and
neutrino-driven convective motions dominate the matter structure behind the shock.
Although each structure of the shock front (grey thin skins in the top panels of Figure~\ref{fig:snap}) and high-entropy bubbles behind the shock is different from the others, 
several large blobs commonly dominate the matter structure behind the shock. 
The shock waves of the low-mass models (s9 and s11) are expanding in a specific direction as shown in the bottom panels of Figure~\ref{fig:snap}. 
On the other hand, the s20 model presents a bipolar-like explosion. This is driven by the matter distribution behind the shock with expanding high-entropy bubbles fragmented by a ``neck'' around the central proto-neutron star, where low-entropy down streams exist.

The shock radius at the end of our simulations is more than several thousands of kilometres, as seen from the box size indicated in the bottom-right corner in each plot of Figure~\ref{fig:snap}. 
This is also the case for the rest of the models (see Figure \ref{fig:snap-all}), and we find that all models examined in this study present successful shock revival. 
This differs from the 3D CCSN study by \citet{burrows20}
who employed the same progenitor series from \citet{sukhbold16} and found that the models s13, s14, and s15 failed in shock revival. 
Cognizant of the influence of magnetic fields, our computations diverge from Burrows' in this aspect. The impact of the weak magnetic field strength in our simulations ($B_0 = 10^{10}$\,G) on the dynamics is, however, very weak \citep[Section~\ref{sec:mag}, see also][]{matsumoto22}.
Differences in numerical schemes for solving the neutrino transport, microphysics, and spatial resolutions, could potentially account for the variance in this shock behavior. 
We note that our simulations show a delayed shock revival for these models (s13, s14, and s15). 
This is qualitatively consistent with the Burrows' results in the sense that these models possess a tendency towards low propensity for explosion. We discuss the reason in the next section.

\subsection{Shock evolution}\label{sec:shock}

\begin{figure}
    \begin{center}
    \includegraphics[width=0.95\linewidth]{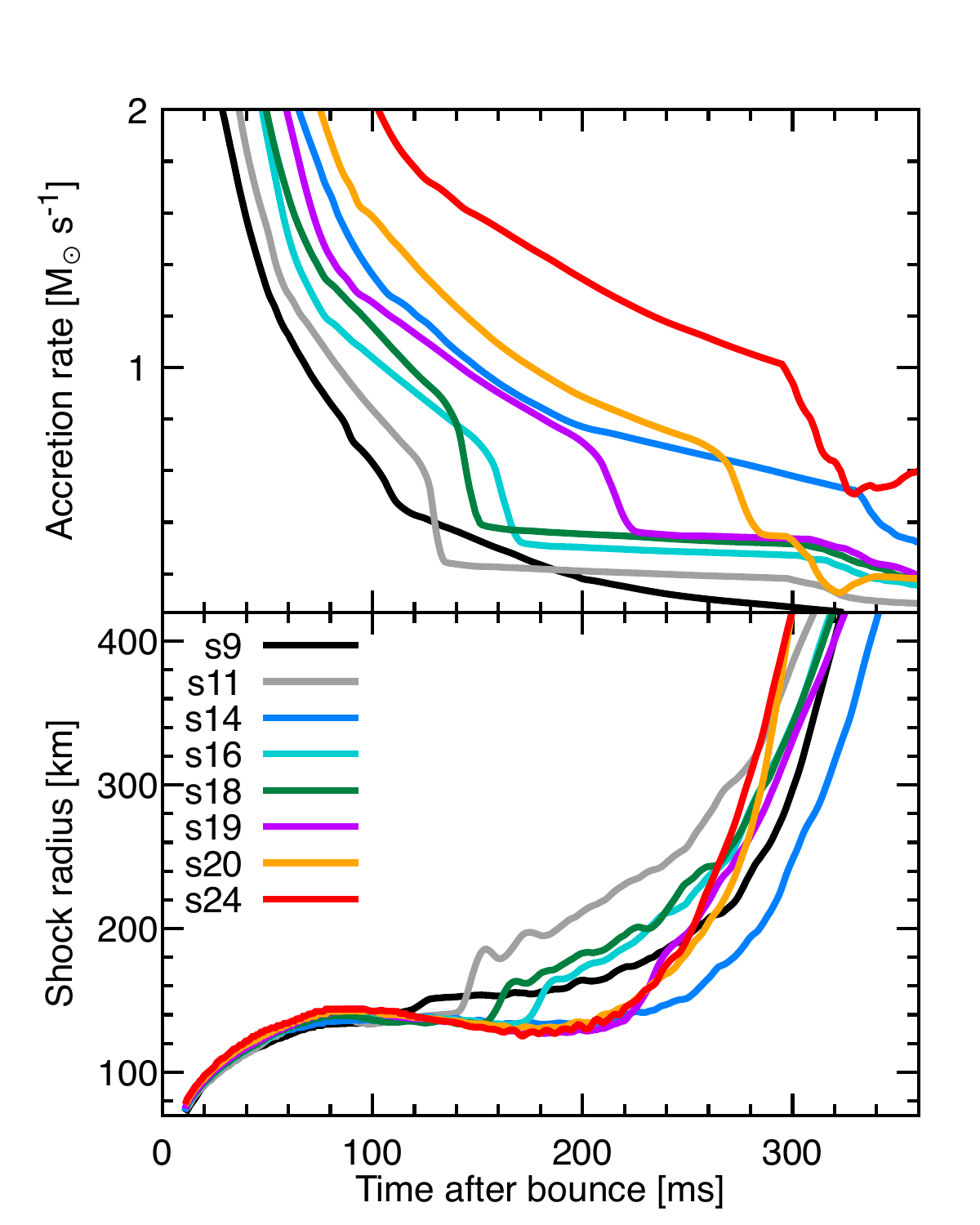}
    \end{center}
    \caption{
    Time evolution of the mass accretion rate at 500\,km in radius (top panel) and the angle-averaged shock radius (bottom panel) in the early 360\,ms. 
    Most of the models show a drop in the accretion rate and a corresponding jump in the shock radius when the Si/O interface falls onto the central region (Figure \ref{fig:prog-den}). 
    The s14 model (blue line) does not hold a sharp density decline at the Si/O interface and it takes a longer time for the shock to turn to the expansion.
    Note that the drops of the accretion rate later than 300 ms after bounce are caused by the expanding shock waves.
    }
    \label{fig:mdot-rsh}
\end{figure}

The bottom panel of Figure~\ref{fig:mdot-rsh} shows the time evolution of angle-averaged shock radii of our models. 
The shock wave of some models (s11, s18, and s16, in order of the timeline) presents a sudden jump up at 140--220\,ms after bounce. These jumps are caused by a density drop structure close to the progenitor centre (Figure~\ref{fig:prog-den}), leading to a decrease in the mass accretion rate shown in the top panel of Figure~\ref{fig:mdot-rsh}. 
As a result, the ram pressure of the matter accreting to the shock wave decreases, and the shock wave turns to expand. 
The s9 model presents an early and small jump of the shock radius at $\sim 120$\,ms, reflecting its small density drop near the central core.
The model s19 also shows such a sudden shock expansion caused by the density drop, but it has to wait until $\sim 230$\,ms after bounce because the density drop is located far from the centre. 
The models s20 and s24 have density drops, but they are farther away from the centre than the model s19, and their shock gradually expands before the density drops fall onto the shock front. 
Despite its intermediate mass, the s14 model demonstrates the slowest shock expansion among the models shown in Figure \ref{fig:mdot-rsh}. The progenitor of the s14 model does not hold a sharp density decline at the Si/O interface and its mass accretion rate is sustained at an elevated level. 
This is also the case for the s13 model (see Figures \ref{fig:prog-den-all} and \ref{fig:mdot-rsh-all}) and 
they take a longer time to turn their stalling shock wave outward. 
We conclude that the magnitude and location of the density jump plays a crucial role in the shock expansion. 

\begin{figure}
    \begin{center}
    \includegraphics[width=0.98\linewidth]{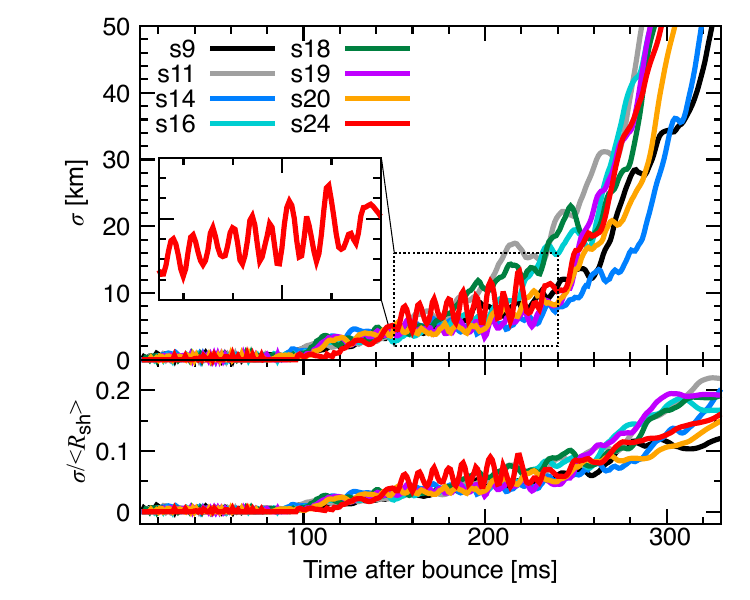}
    \end{center}
    \caption{Rms deviation $\sigma$ of the shock radius from its angle-averaged value. The inset in the top panel extracts the $\sigma$ of the s24 model in 150--240\,ms post bounce. It presents a quasi-periodic oscillation in this phase, corresponding to the spiral SASI motion of the shock front. The ratio of the $\sigma$ to the angle-averaged shock radius (bottom panel) is less than 10 percent before shock revival.}
    \label{fig:sigma}
\end{figure}

In typical multi-dimensional CCSN models with neither extreme rotation nor strong magnetic fields, successful explosions are achieved by neutrino heating assisted by hydrodynamic instabilities such as convective motion \citep[increasing the dwelling time of shocked material within the gain region, see ][]{Foglizzo06} and the Standing Accretion Shock Instability (SASI; \citealt{Blondin03}; \citealt{Foglizzo07}). 
An intense sloshing mode of SASI boosts the neutrino-heating mechanism by developing large-scale oscillations of the shock front and spreading out the gain region where neutrinos deposit energy into the matter. 
To assess the power of the SASI sloshing motion, we estimate a root-mean-squared (rms) deviation $\sigma$ of 
the angle-dependent shock radius $R_{\rm sh}(\theta, \phi)$ 
from its angle-averaged value $\langle R_{\rm sh} \rangle$, 
defined by 
\begin{equation}
    \sigma = \sqrt{\int (R_{\rm sh}(\theta, \phi) - \langle {R_{\rm sh} \rangle)^2 d\Omega}/(4 \pi)}.
\end{equation}
The time evolution of $\sigma$ before and after the shock revival is shown in Figure~\ref{fig:sigma}. 
The shock deformation level during the shock stalling phase ($< 200$\,ms after bounce) is a few percent 
for all models and any clear SASI sloshing motions do not appear in our models. 
The model s24 presents a quasi-periodic oscillation in this phase, although its amplitude is small. 
This feature is caused by a spiral mode of SASI, which is discussed later. 

\begin{figure}
    \begin{center}
    \includegraphics[width=0.98\linewidth]{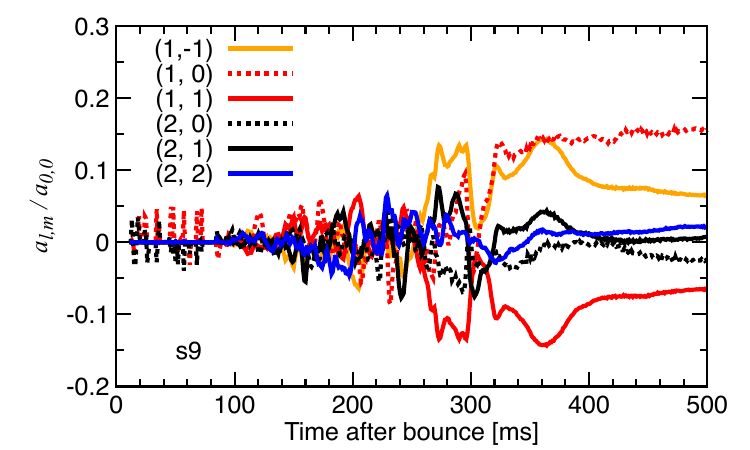}
    \includegraphics[width=0.98\linewidth]{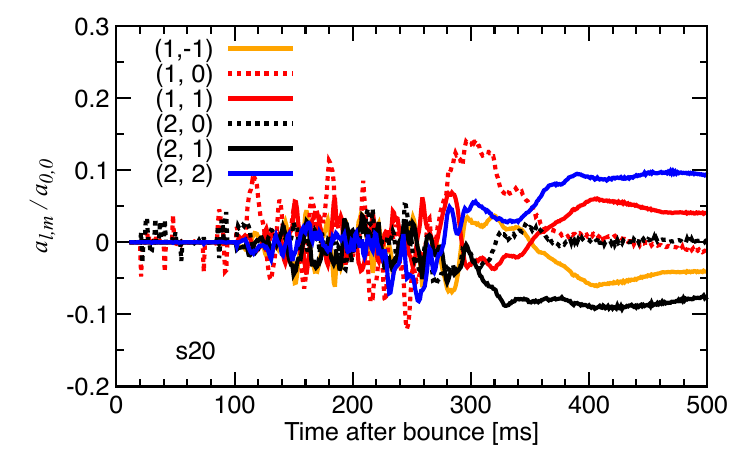}
    \includegraphics[width=0.98\linewidth]{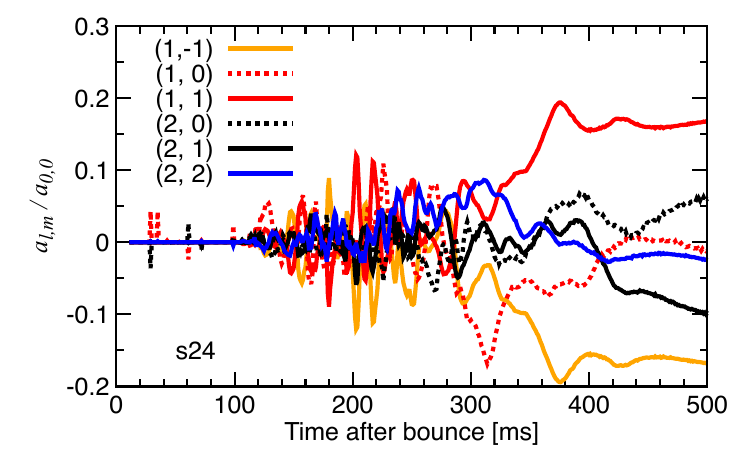}
    \end{center}
    \caption{Time evolution of decomposition factors $a_{l,m}$ of the angle-dependent shock radius. 
    At $\tpb = 500$\,ms, the dominant modes of the model s9 (top panel), and the model s20 (middle panel), are $l=1$ (orange and red lines), and $l=2$ (black and blue lines), respectively. These features correspond to the shock morphology found in Figure \ref{fig:snap}, that is, a unipolar-like explosion of s9 and a neck-choked bipolar-like explosion of s20.
    The model s24 (bottom panel) shows a development of the $(l,m)=(1,\pm 1)$ modes (solid orange and red lines) around $\tpb = 200$\,ms, which could be an expression of the spiral SASI mode (see also Figure \ref{fig:sigma}).}
    \label{fig:alm}
\end{figure}

We also evaluate decomposition factors $a_{l,m}$ of the angle-dependent shock radius. 
Figure~\ref{fig:alm} shows $l=1$ and $2$ modes of the models s9, s20, and s24.
For the s9 model, all modes are weak before shock revival, and $l=1$ modes (orange and red lines) become dominant after shock revival ($t_{\rm pb} > 300$\,ms). 
On the other hand, the model s20 presents a relatively large amplitude of $(l,m)=(1,0)$ mode (red dotted line) even before the shock revival. 
Although the time oscillation of this mode is a character of the SASI sloshing mode, it is not enough to globally deform the shock front (Figure~\ref{fig:mdot-rsh}). 
Accompanied by the small $\sigma$ before the shock revival (Figure~\ref{fig:sigma}), we conclude that these models do not show SASI sloshing motions. 
In contrast to the model s9, the shock decomposition factor $a_{l,m}$ of the model s20 after the shock revival is dominated by $l=2$ modes (black and blue lines), as visualized by the dipolar-like profile in Figure~\ref{fig:snap}. 
In contrast to the models s9 and s20, the s24 model shows a development of $(l,m)=(1,\pm 1)$ mode (orange and red solid lines) around 200 ms after bounce. This is construed as corroboration of a spiral SASI motion appeared in the s24 (and s23, shown in Figure \ref{fig:sigma-all}) model.

\subsection{Energetics}\label{sec:energy}
\begin{figure}
    \begin{center}
    \includegraphics[width=0.95\linewidth]{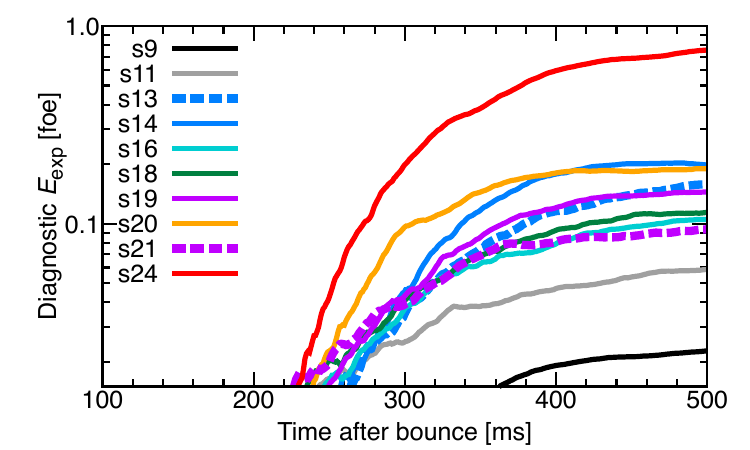}
    \end{center}
    \caption{Time evolution of the diagnostic explosion energy. 
    Among the examined models in this study, the lightest model s9 shows a small explosion energy, while the heaviest model s24 demonstrates a large explosion energy; however, the intermediate models exhibited a somewhat non-monotonic tendency.
    See Figure \ref{fig:eexp-all} for the corresponding plot with full model sets.}
    \label{fig:eexp}
\end{figure}

Current understanding of the standard mechanism of CCSN explosion is that 
the explosion is powered by neutrino heating, and its energy source is the gravitational potential released by a collapsing core and the matter accreting onto it.
Therefore, 
the final explosion energy of the models s9, s11, s16 and s18, which have a small energy supply from the accreting matter, is relatively small ($\sim 10^{50}$\,erg or less) among the examined models despite the early shock revival. 
Here, we define the diagnostic explosion energy $E_{\rm exp}$ as 
\begin{equation}\label{eq:eexp}
E_{\rm exp} = \int_D \left( \frac{1}{2}\rho \boldsymbol{v} \cdot \boldsymbol{v} - \rho \Phi + e_{\rm int} + \frac{1}{8 \pi} \boldsymbol{B} \cdot \boldsymbol{B} \right).
\end{equation}
The integrand is the sum of kinetic, gravitational, internal, and magnetic energy 
and the integrating region $D$ represents the domain where the integrand is positive and the materials have positive radial velocity. 
The time evolution of $E_{\rm exp}$ is shown in Figure~\ref{fig:eexp}. 
The lightest progenitor among our models, s9, shows a small explosion energy ($\sim 0.02 \times 10^{51}$\,erg at 0.5 s after bounce). 
The most energetic explosion is obtained by the s24 model ($\sim 0.8 \times 10^{51}$\,erg), which is compatible with the typical explosion energy of observed CCSNe.

Figure~\ref{fig:eexp}, as well as Figure~\ref{fig:eexp-all} in the Appendix, clearly indicate that the explosion energy is not arranged in order of progenitor’s ZAMS mass. 
For example, the explosion energy of the models s13 and s14 is as high as that of the models s19 and s20, whereas the s21 model shows a relatively small explosion energy. 
This is because the mass accretion, which eventually drives the explosion, exhibits a non-monotonic relationship with the ZAMS mass. 
Since the progenitors s13 and s14 have a small or negligible density drop at the Si/O interface, they maintain a high accretion rate for an extended period (Figure~\ref{fig:mdot-rsh-all}). 
Quantities characterizing the accretion rate, such as the compactness parameter, should well correlate with the explosion energy once the final converged values of the explosion energy are determined by longer calculations.

\section{Protoneutron star}\label{sec:pns}

As described in the previous section, all models examined in this study demonstrate successful shock revival within 300 ms after bounce. 
It suppresses further matter accretion onto the central cores, resulting in the compact remnants being neutron stars (NSs), not black holes. 
This section examines the properties of the (proto-)NS such as mass, spin, and kick velocity.

\begin{figure*}
    \begin{center}
    \begin{tabular}{cc}
    \includegraphics[width=0.46\linewidth]{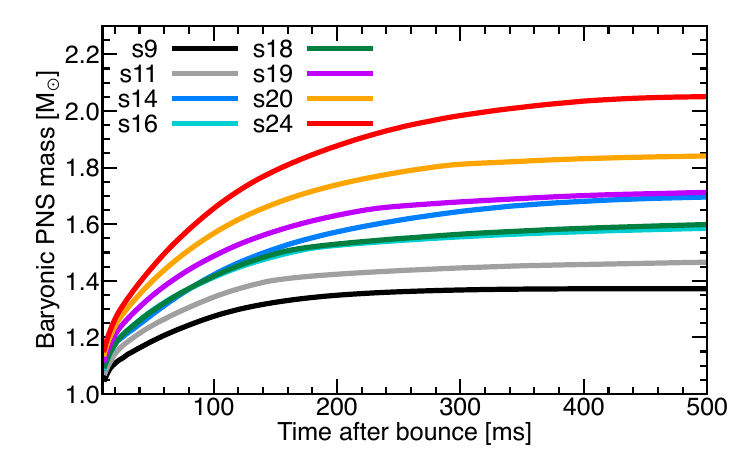} &
    \includegraphics[width=0.46\linewidth]{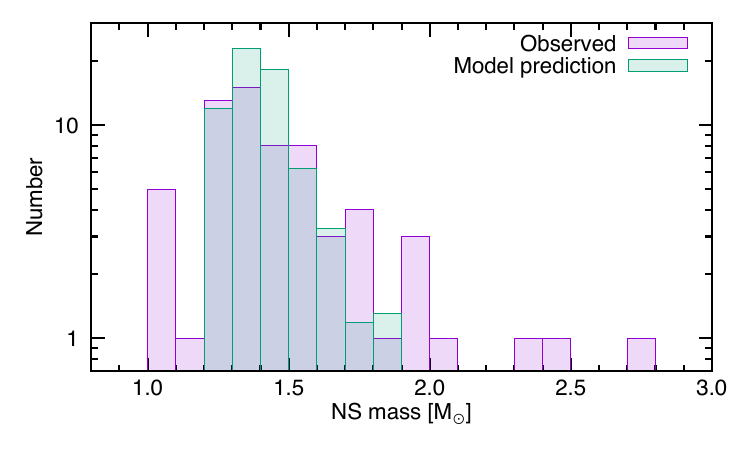} \\
    \end{tabular}
    \end{center}
    \caption{Mass properties of the central remnants. 
    Left panel: time evolution of the baryonic PNS mass. 
    Right panel: IMF-weighted final gravitational mass distribution (green) compared with the observations (purple). The data is taken from  \citet{lattimer12}.}
    \label{fig:pns-mass}
\end{figure*}

We define the surface of NS at the radius where mass density $\rho$ equals $10^{11}$\,g\,cm$^{-3}$. 
The left panel of Figure~\ref{fig:pns-mass} shows the time evolution of the baryonic PNS mass. 
Models with a high mass accretion rate, characterized by high compactness $\xi_M$, produce massive PNS.
The PNS mass is almost saturated within our simulation time and, at the final simulation time ($t_{\rm pb} = 0.5$\,s), it ranges from $1.37\,\Msun$ (s9) to $2.08\,\Msun$ (s23). 

We estimate the gravitational mass of the final cold neutron star and compare an IMF-weighted number distribution with observation (the right panel of Figure~\ref{fig:pns-mass}). 
We define the gravitational NS mass as the rest mass, $\int \rho \, dV$, minus the mass equivalent of the binding energy, $\frac{1}{2} \int \rho \Phi \, dV$.
Here, we assume Salpeter IMF and the observed data is from Table~1 in \citet{lattimer12}, see also \cite{Ozel16,fonseca21,Badry24} for more recent catalog and findings. 
We find that our model prediction for the NS mass is peaked at $1.4\,\Msun$, and the majority is in the range of $1.2\,\Msun < M_{\rm NS} < 1.7\,\Msun$. 
These features are consistent with observations, except very low mass ($\sim 1.0\,\Msun$) and high mass ($>2\,\Msun$) NSs, see \citet{suwa18,muller24,salmi24} for the discussion. 
The anomalous mass of these NSs could potentially be attributed to binary interactions not considered in this paper. 
Low-mass NSs might come from ultra-stripped SN explosions that have undergone mass stripping via binary interaction. 
High mass NSs might be 
the products of mass feeding in binary systems and 
supported by rapid rotation. 

\begin{figure*}
    \begin{center}
    \begin{tabular}{cc}
    \includegraphics[width=0.46\linewidth]{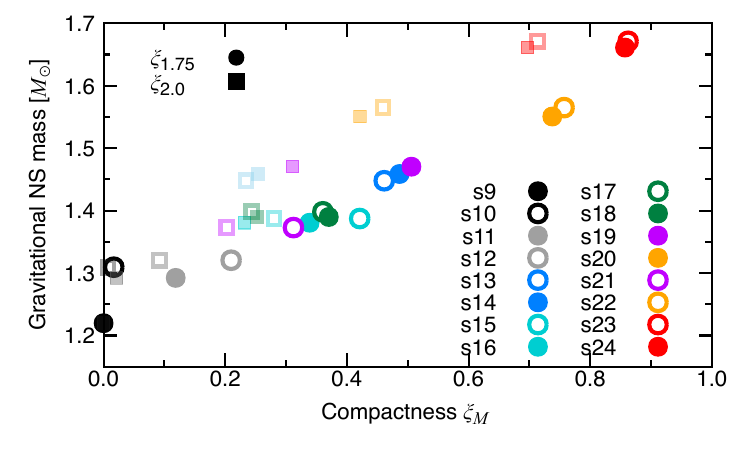} &
    \includegraphics[width=0.46\linewidth]{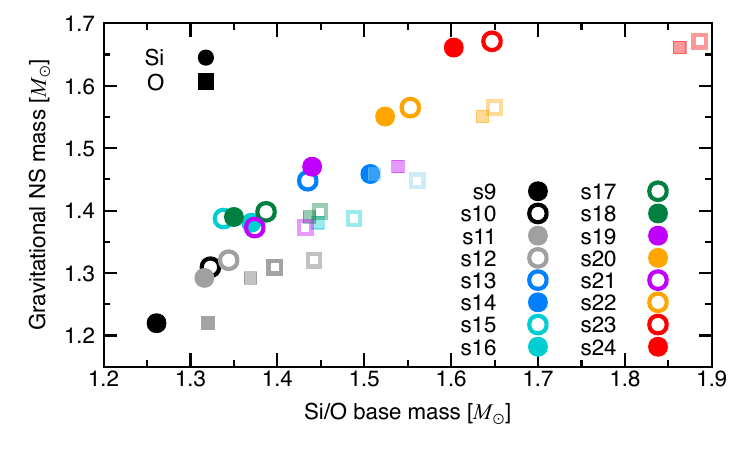} \\
    \end{tabular}
    \end{center}
    \caption{Left panel: relation between the gravitational neutron star mass and the compactness parameters $\xi_M$ at $M=1.75\,\Msun$ (circles) and $2.0\,\Msun$ (squares).
    Right panel: relation between the gravitational neutron star mass and the base mass of the Si layer (circles) and the O layer (squares). See Figure \ref{fig:prog} for the definition of the base mass of the Si/O layers.}
    \label{fig:pns-mass2}
\end{figure*}

As seen from the left panel of Figure \ref{fig:pns-mass} and the top-right panel of Figure \ref{fig:eexp-all}, the mass of NSs is almost convergent within our simulation time. The time evolution of the NS mass is dominated by the mass accretion rate, which can be well characterized by the compactness parameter $\xi_M$. Thus, the relation between the final NS mass and the compactness parameter shows an approximately monotonic behavior (the left panel of Figure \ref{fig:pns-mass2}). This behavior has been found in our previous 2D study \citep{nakamura15} and here confirmed by 3D simulations. 
A similar monotonic relation is found between the NS mass and the base mass of the Si and O layers (the right panel of Figure \ref{fig:pns-mass2}). These correlations suggest the possibility of predicting one of the final evolutionary states of CCSNe from the initial progenitor structure.

\begin{figure*}
    \begin{center}
    \begin{tabular}{cc}
    \includegraphics[width=0.46\linewidth]{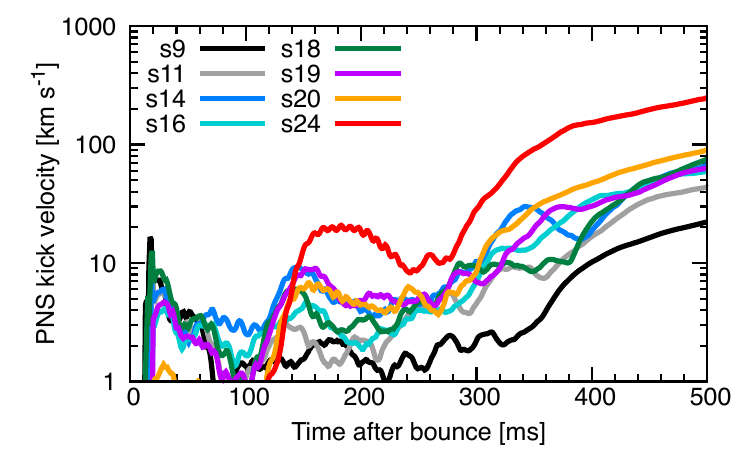} &
    \includegraphics[width=0.46\linewidth]{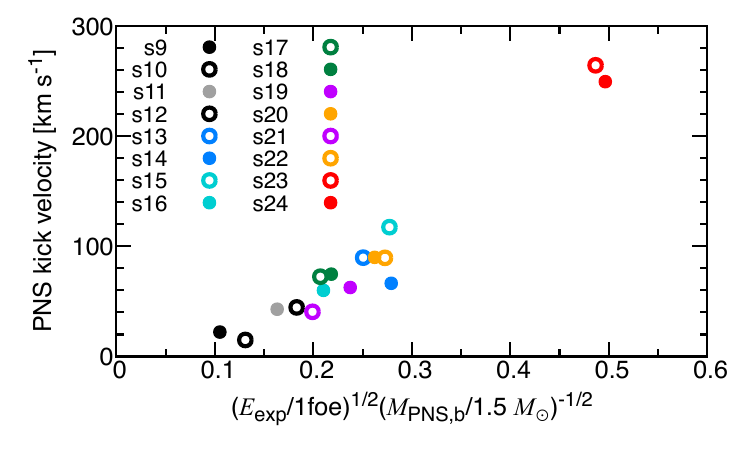}
    \end{tabular}
    \end{center}
    \caption{Left panel: time evolution of the PNS kick velocity. At the final time of our simulations, the kick velocity is mainly in the range $\sim$ 20--100 km s$^{-1}$, whereas the fastest one is $\sim 250$ km s$^{-1}$.
    To improve readability, the lines have each been smoothed using a 20-ms moving average. 
    Right panel: relation between the PNS kick velocity and the quantity $(E_{\rm exp}/M_{\rm PNS,b})^{1/2}$, where $E_{\rm exp}$ is the explosion energy and $M_{\rm PNS,b}$ is the baryon mass of the PNS at the final time of our simulations.}
    \label{fig:kick}
\end{figure*}

As described in Figures~\ref{fig:snap} and~\ref{fig:alm}, our CCSN models present highly asymmetric shock expansion. 
This aspherical explosion causes a recoil (hydrodynamic ``kick'') of NSs. 
We estimate the kick velocity in a slightly different manner from 
the previous 2D studies \citep{scheck04,nakamura19}.
We assume momentum conservation within the computational domain and estimate the PNS kick velocity, $v_{\rm kick}$, by the following formula:
\begin{equation}
    v_{\rm kick} = \left| \boldsymbol{P}_{\rm gas} \right| / M_{\rm PNS}
    = \left| \int_{R>R_{\rm PNS}} \rho \boldsymbol{v} \, dV \right| / M_{\rm PNS},
\end{equation}
where $\boldsymbol{P}_{\rm gas}$ is the total momentum of the gas outside of the PNS, with the fluid density $\rho$ and velocity $\boldsymbol{v}$ at each time.
We find that the PNSs are accelerated after the shock revival, and the final values at $t_{\rm pb} = 0.5$\,s range from $\sim 20 \,{\rm km\,s}^{-1}$ (s9) to $\sim 250 \,{\rm km\,s}^{-1}$ (s24) 
(Figure~\ref{fig:kick}). 
As discussed in \citet{nakamura19}, $v_{\rm kick}$ is predominantly determined by the strength of the explosion, which can be characterized by the explosion energy. 
Although matter expansion weighted in one direction, as seen in the models s9 and s11 (Figure~\ref{fig:snap}), is preferable for PNS acceleration than a bipolar-like explosion (s20), the weak explosion of these models results in small $v_{\rm kick}$. 
To clearly show this point, we plot the PNS kick velocity as a function of the quantity $(E_{\rm exp}/M_{\rm PNS,b})^{1/2}$ (the right panel of Figure~\ref{fig:kick}), where $E_{\rm exp}$ is the explosion energy and $M_{\rm PNS,b}$ is the baryon mass of the PNS, at 0.5 s after bounce, following \citet{burrows24}.

Asymmetric neutrino radiation also produces the NS kick \citep{nagakura19_kick}. 
Our 3D MHD CCSN models show asymmetrical neutrino fluxes and are expected to have a kick velocity component coming from the neutrino radiation asymmetry, in addition to the component from hydrodynamic recoil. 
In the forthcoming paper, we will estimate multi-messenger signals from our CCSN models.

\begin{figure}
    \begin{center}
    \includegraphics[width=0.95\linewidth]{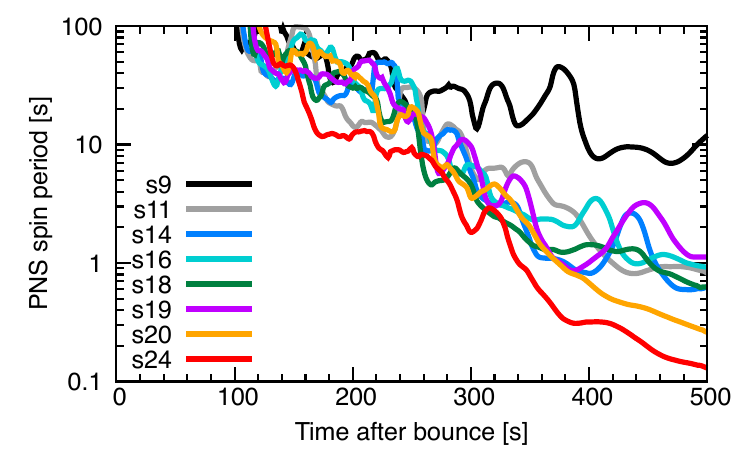}
    \end{center}
    \caption{Time evolution of the PNS spin period. 
    The lines are smoothed as in Figure \ref{fig:kick}.}
    \label{fig:spin}
\end{figure}

\begin{figure}
    \begin{center}
    \includegraphics[width=0.95\linewidth]{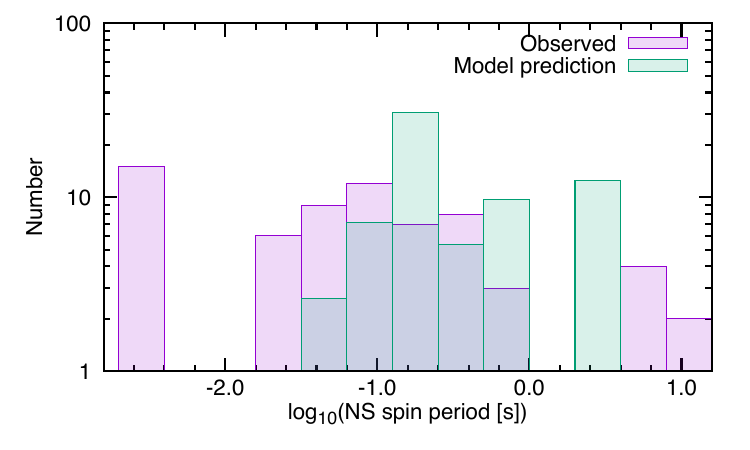}
    \end{center}
    \caption{PNS spin period distribution predicted by our models (green) compared with the observations (purple). The observation data is taken from \citet{Igoshev22}.}
    \label{fig:spindist}
\end{figure}

Although our simulations start from spherically symmetric matter distribution, non-radial fluid motions develop via hydrodynamic instabilities, partially with the aid of the non-spherical magnetic fields. 
A fraction of the matter accumulates in the central region and spins up the PNS. 
We estimate the angular momentum of the PNS $\boldsymbol{J}$ in the same manner as in \cite{nakamura22}: 
\begin{equation}\label{eq:ang_mom}
    \frac{d\boldsymbol{J}}{dt} = \int_{r=100{\rm km}} \rho v_r \boldsymbol{v} \times \boldsymbol{r} \,d\Omega.
\end{equation}
A moment of inertia, $I$, of the NS is given by
\begin{equation}
I = 0.237 M_{\rm grav} R^2
 \left[ 1+4.2 \left( \frac{M_{\rm grav} / \Msun}{R /{\rm km}} \right) 
+ 90  \left( \frac{M_{\rm grav} / \Msun}{R /{\rm km}} \right)^4 \right].\label{eq:mom_inrt}
\end{equation}
Then, we obtain a rough estimate of the NS spin period $T_{\rm NS}$ from $T_{\rm NS} = 2 \pi \, I / |\boldsymbol{J}|$. 
Figure~\ref{fig:spin} shows the time evolution of $T_{\rm NS}$. 
Our estimations show that 
the NSs rotates with a period of $\sim 10$\,s (s9) -- $\sim 0.1$\,s (s24) at 0.5\,s after bounce, 
eventually spinning up to 3.8\,s (s9) -- 0.04\,s (s24), 
assuming the final NS radius to be 12\,km and the final NS mass to be the value at 0.5\,s post bounce.
Figure \ref{fig:spindist} compares the distributions of the NS spin rate between our model prediction and observations \citep{Igoshev22}. 
Our models show a peak at $T_{\rm NS} \sim$0.1--0.2\,s, roughly matching the observations.

\begin{figure}
    \begin{center}
    \includegraphics[width=0.95\linewidth]{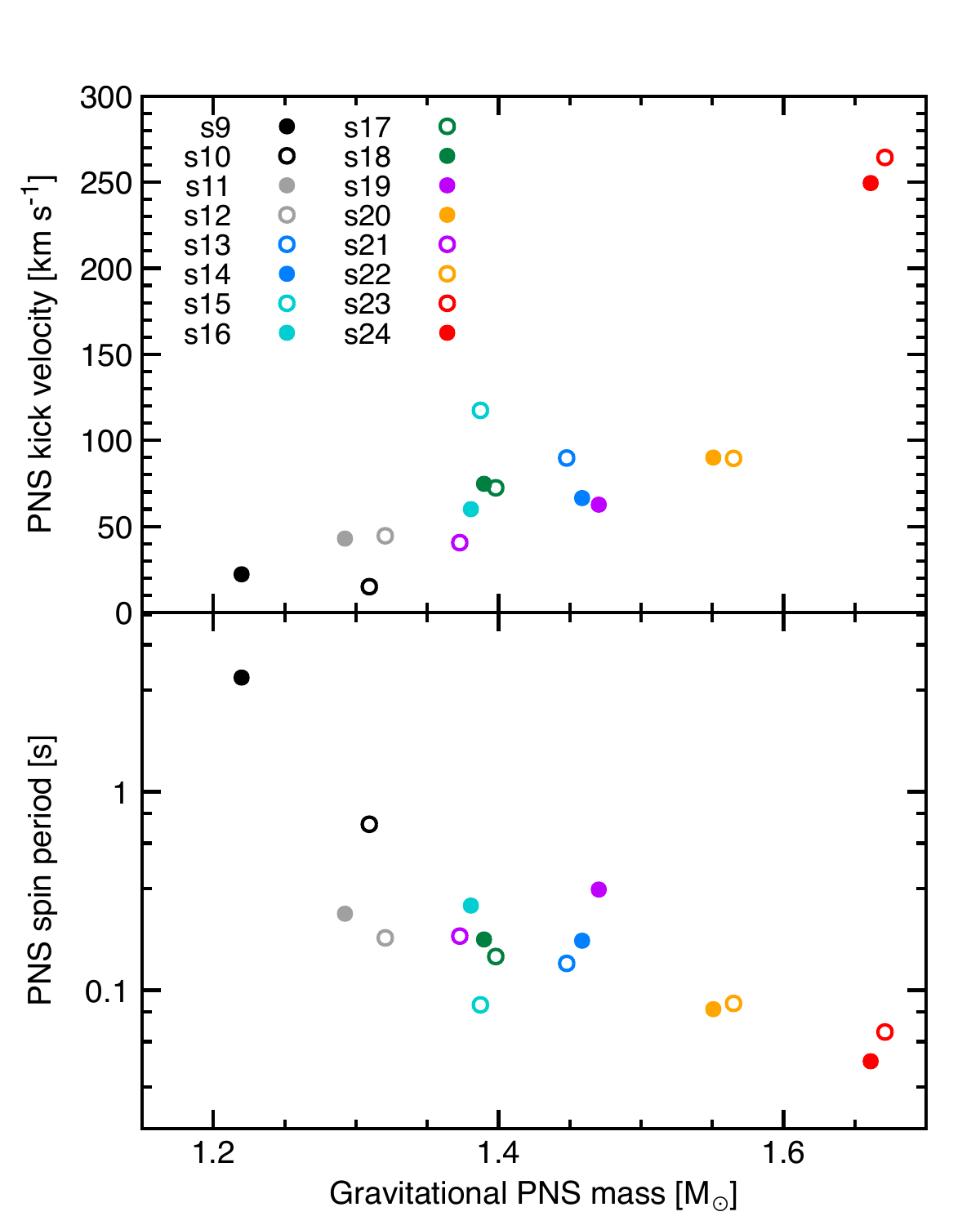}
    \end{center}
    \caption{PNS kick velocity (top panel) and spin period (bottom panel) as a function of the gravitational PNS mass.}
    \label{fig:mpns-pns}
\end{figure}

Interestingly, our mdoels predict that 
the NS spin periods and kick velocities are roughly in the same order and well (anti-)correlates with the explosion energy. 
It is reasonable that energetic explosion produces a large kick velocity. 
The apparent anti-correlation between the explosion energy and the NS spin period arises from their mutual dependence on the mass accretion onto the central region including PNS. 
The energy source of neutrinos that trigger neutrino-driven explosions is the gravitational potential energy of matter falling inward. 
A high accretion rate is synonymous with a high heating rate, and therefore, the explosion energy of models with high accretion rates increases. 
Angular momentum is also accumulated in the PNS via the matter accretion. 
Therefore, the key factor laying behind these (anti-)correlations is the mass accretion, which is characterized by the compactness parameter $\xi_M$. 
In Figure \ref{fig:mpns-pns} we plot the NS kick velocity and the spin period against the gravitational mass of the PNS instead of $E_{\rm exp}$ or $\xi_M$, 
since the NS mass is directly observable and the increase in the mass is the most direct consequence of mass accretion.

\begin{table*}
\caption{Properties of explosions and PNSs}
\begin{tabular}{rccccccccc}
\hline
model & $t_{400}$ & $R_{\rm sh}$ & $E_{\rm dia}$ & $M_{\rm PNS,b}$ & $M_{\rm PNS,g}$ & $R_{\rm PNS}$& $v_{\rm kick}$ & $T_{\rm NS}$ & $E_{\rm PNS,mag}$ \\
& [s] & [$10^3$km] & [foe] & [$\Msun$] & [$\Msun$] & [km] & [km\,s$^{-1}$] & [s] & [$10^{46}$erg] \\
\hline
 s9& 0.320 &  2.42 &  0.02 &  1.37 &  1.22 & 23.4 &  22 &  3.76&   0.3 \\
s10& 0.385 &  1.25 &  0.04 &  1.49 &  1.31 & 24.0 &  15 &  0.69&  37.9 \\
s11& 0.304 &  1.83 &  0.06 &  1.47 &  1.29 & 24.3 &  42 &  0.24&  19.5 \\
s12& 0.336 &  1.46 &  0.08 &  1.51 &  1.32 & 23.7 &  44 &  0.18& 159.8 \\
s13& 0.342 &  1.58 &  0.16 &  1.68 &  1.45 & 25.2 &  89 &  0.14&   7.3 \\
s14& 0.338 &  1.95 &  0.20 &  1.70 &  1.46 & 25.1 &  66 &  0.18&  61.1 \\
s15& 0.334 &  1.82 &  0.18 &  1.60 &  1.39 & 24.7 & 117 &  0.08&  51.0 \\
s16& 0.314 &  1.74 &  0.11 &  1.59 &  1.38 & 24.6 &  60 &  0.27&  33.4 \\
s17& 0.313 &  1.91 &  0.10 &  1.61 &  1.40 & 24.8 &  72 &  0.15&  26.2 \\
s18& 0.315 &  1.74 &  0.11 &  1.60 &  1.39 & 24.8 &  74 &  0.18&  31.5 \\
s19& 0.320 &  1.88 &  0.14 &  1.71 &  1.47 & 24.9 &  62 &  0.32&  25.4 \\
s20& 0.298 &  2.41 &  0.19 &  1.84 &  1.55 & 23.9 &  90 &  0.08&  52.3 \\
s21& 0.300 &  1.90 &  0.09 &  1.57 &  1.37 & 24.9 &  40 &  0.19&  15.0 \\
s22& 0.295 &  2.27 &  0.21 &  1.87 &  1.56 & 23.9 &  89 &  0.09&  20.6 \\
s23& 0.293 &  2.94 &  0.74 &  2.08 &  1.67 & 22.8 & 264 &  0.06&  64.2 \\
s24& 0.297 &  2.83 &  0.76 &  2.05 &  1.66 & 22.9 & 249 &  0.04& 291.9 \\
\hline
\multicolumn{9}{l}{All values except $t_{400}$ and $T_{\rm NS}$ are evaluated at 0.5 s after bounce.}\\
\multicolumn{9}{l}{The surface of the PNS is defined at the radius where the mass density is $10^{11}\,{\rm g\,cm}^{-3}$.}
\label{tbl:final}
\end{tabular}
\end{table*}
% from MBP:Documents/0_research/1_my-papers/oo_3DMHD/mktbl-3d/

\section{Impact of magnetic fields}\label{sec:mag}

\begin{figure}
    \begin{center}
    \includegraphics[width=0.95\linewidth]{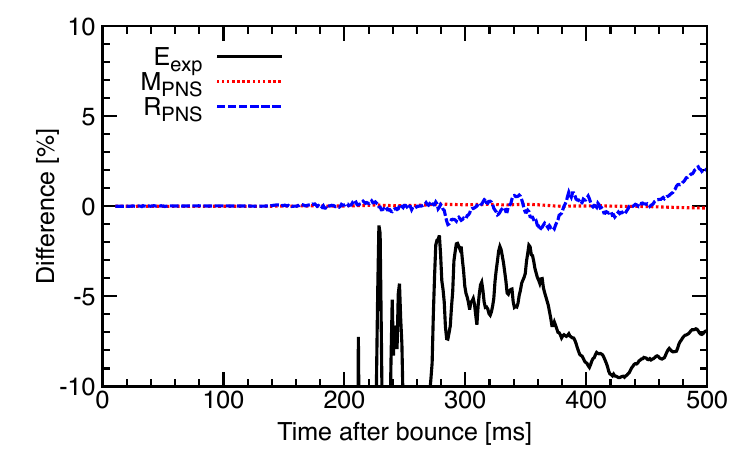}
    \end{center}
    \caption{Comparison between the s23 models with and without the magnetic field. Shown are the difference of some explosion properties $X$ defined by $(X_{\rm no \, mag.} - X_{\rm mag.})/X_{\rm mag.}$.}
    \label{fig:comp-s23}
\end{figure}

Our MHD models of CCSNe assume relatively weak magnetic fields (initially $10^{10}$ G at the centre).
To assess the impact of the magnetic fields on the evolution of hydrodynamic flows and the PNS, we have conducted an additional simulation without magnetic field ($B_0=0$) for a $23\,\Msun$ progenitor 
and found that these two models with and without magnetic fields show very similar evolution (Figure~\ref{fig:comp-s23}). 
At $0.5$\,s after bounce, the difference of explosion properties, such as shock radius and PNS mass and radius, are within a few percent. 
A relatively large difference appears in the explosion energy. 
The model without a magnetic field shows $E_{\rm dia}=0.69$ foe, which is 7 percent smaller than the model with a magnetic field. 
We note that the contribution of the magnetic energy to the total explosion energy remains below 1 percent in our simulation.

As discussed in \citet{matsumoto22}, the magnetic field strength employed in this study plays only a limited role in supernova dynamics, although they used different progenitor models.
Initially strong magnetic fields can be accumulated and enhanced at the surface of neutrino-driven convective motion behind a shock wave, leading to faster and more energetic explosions with the aid of magnetic pressure \citep{matsumoto22}.

\begin{figure*}
    \begin{center}
    \begin{tabular}{cc}
    \includegraphics[width=0.48\linewidth]{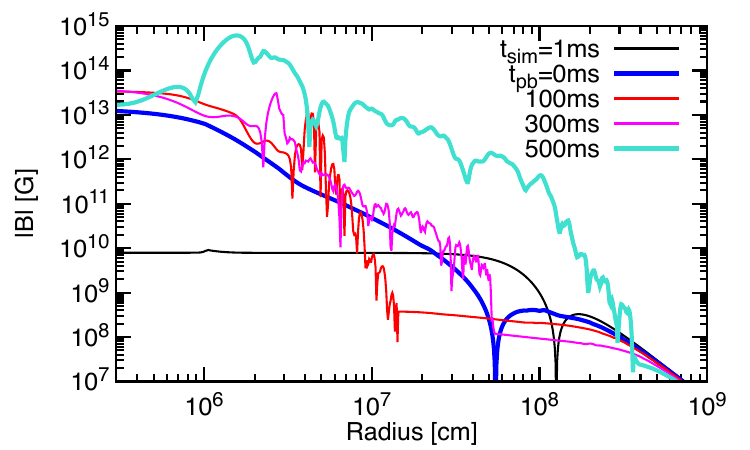} &
    \includegraphics[width=0.48\linewidth]{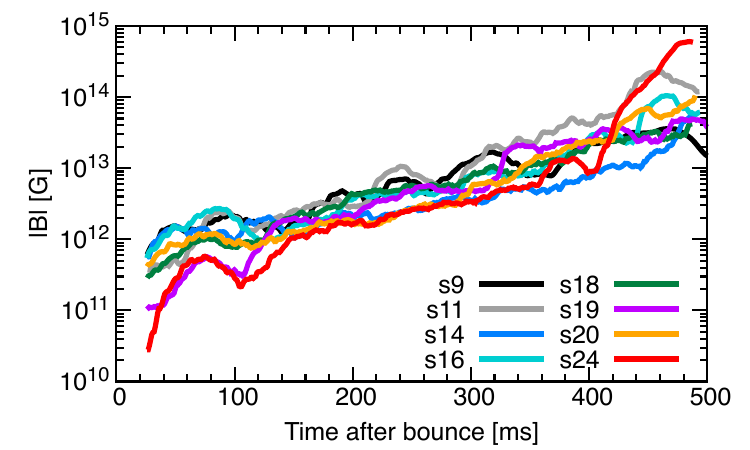} \\
    \end{tabular}
    \end{center}
    \caption{Time evolution of the magnetic field. Left: radial profile of the total magnetic strength for the s24 model for some selected times. Right: time evolution of the total magnetic strength at the PNS surface.
    }
    \label{fig:mag}
\end{figure*}

Although the strength of the initial magnetic fields assumed in our simulations is weak and its impact on the dynamics is limited, the magnetic fields are enhanced during the collapse and post-bounce phases. 
The left panel of Figure \ref{fig:mag} shows the radial profile of the total magnetic strength of the s24 model for some selected times. 
The central magnetic field strength is amplified to $\sim 10^{13}$\,G at the time of core bounce. This amplification is due to the deposition of magnetic fields frozen to accreting materials. 
Then the domain with the amplified magnetic fields expands as the shock wave goes outward. 
The magnetic field is accumulated and amplified to the magnetar level ($\mathcal{O}$($10^{14}$ --$10^{15}$)G) in the convectively stable shell in the vicinity of the PNS surface (20--30 km at this phase) at the final simulation time.
Despite differing initial conditions, \citet{matsumoto22} identified a magnetic field amplification in a similar region, the convectively stable shell in the vicinity of the PNS surface, using the same numerical code. They concluded that a turbulent magnetic diffusion in the PNS convection, as well as the deposition of magnetic fields from the accreting materials, contributes to the magnetic field amplification in the convectively stable region. Although we do not delve into more about the details of the amplification mechanism, it is reasonable to infer that a similar mechanism operates in our models. 
Refer to \citet{matsumoto22} for more details about the amplification process. 
The right panel of Figure \ref{fig:mag} presents the time evolution of the total magnetic strength at the PNS surface for our models. 
It can be seen that the final magnetic strength strongly depends on the models. 
The outstanding enhancement of the magnetic field of the model s24 is not directly correlated to its fast PNS spin, since the non-radial motion of the core is artificially suppressed in our simulations and the spin period shown in Figure \ref{fig:spin} is estimated by integrating the angular momentum of the accreting material. The potential possibility for spontaneous rotational motion to drive magnetic field amplification, however, presents an intriguing subject of further investigation.

\section{Summary and discussions}\label{sec:summary}

In this paper, 
we report the hydrodynamic evolution of our 3D CCSN models and the properties of their central remnants. 
We employ sixteen progenitor models from \citet{sukhbold16}, covering a ZAMS mass range from $9 \, \Msun$, almost the lower limit of Fe-core collapse SN, to $24 \, \Msun$, including a peak of the compactness parameter in this mass range. 
Adding a fixed magnetic field structure to the non-magnetized progenitor models as an initial condition, we have conducted 3D MHD simulations from the core collapse up to 0.5\,s after bounce. 
The strength and distribution of the magnetic fields in CCSN cores are not known and might be dependent on the progenitor stars. 
Focusing on a systematic study on the progenitor's matter structure dependence of the supernovae, 
we fix the initial magnetic field to be a dipole-like distribution with the central strength of $10^{10}$ G 
and cutoff at 1000 km for all progenitors. 
Note that the weak magnetic field assumed in our models does not have any impact on dynamics \citep[Section~\ref{sec:mag}, see also][]{matsumoto20,matsumoto22}.

Although our models are limited in time until 0.5 s after bounce, 
we have found that in all examined models the shock waves once stall at $\sim 150$\,km and finally turn to expansion. 
The time evolution of the shock radius strongly correlates to the mass accretion history. 
Most of the progenitor models possess a density drop at the Si/O interface. 
When it falls onto the shock front, a sudden decrease of the ram pressure from the accreting matter causes the stalling shock wave to pop. 
This popping increases the volume of the gain region behind the shock, resulting in a runaway expansion of the shock wave. 
\citet{burrows20} reported that their 3D simulations starting from $13\,\Msun$, $14\,\Msun$, and $15\,\Msun$ progenitors from \citet{sukhbold16} failed in explosion. 
The $13\,\Msun$ and $14\,\Msun$ models (s13 and s14) have little density jump at the Si/O interface, and we find that these models do not experience (s14) or experience a tiny drop (s13, at 260 ms after bounce) in the mass accretion rate.
As a result, they take relatively more prolonged time among our models to revive the shock wave.
The s15 model is in a different situation. It has a small but clear density jump structure at 1,500 km in radius and a corresponding decrease in the mass accretion rate at 90 ms after bounce (Figures \ref{fig:prog-den-all} and \ref{fig:mdot-rsh-all}). 
The location of the density jump is close to the center, so that the accretion rate is sustained at high level during the shock stalling phase, which might causes the failed explosion in Burrows' model. 
The difference between our exploding models and the failed models in \citet{burrows20} might be caused by different numerical treatments such as neutrino transport scheme, applied microphysics and equation of state, and spatial resolution, as well as including/not-including the effects of magnetic fields, although the weak magnetic strength assumed in our simulations has little impact on the dynamics. 

The compactness parameter, which is a very simple parameter defined by \citet{Oconnor11a} as $\xi_M = M/R$ and widely used to analyze CCSN properties, can characterize the global structure of the density profile, 
whereas it can not capture small-scale structures like the density drop at the interface of chemical layers. 
We confirm that the density drop plays a crucial role in the shock revival; in this sense, the compactness parameter is unsuitable for determining the explodability. 
On the other hand, the explosion energy is well correlated to the compactness. 
Since the energy source of the neutrinos which heats the matter behind the shock is the gravitational potential energy of the accreting matter, 
the models with high accretion rate, which  means high compactness, present high explosion energy if they explode.
In fact, the most powerful explosions in our study are obtained by $23\,\Msun$ and $24\,\Msun$ models, locating at the peak of the compactness parameter distribution in the mass range examined in this paper. 
Their final value of the diagnostic explosion energy, $E_{\rm dia} \sim 0.75 \times 10^{51}$ erg, is close to but less than the ``standard'' explosion energy ($10^{51}$ erg). 
A recent supernova survey project, however, suggests a smaller explosion energy for the typical (median) value \citep[$\sim 0.6 \times 10^{51}$ erg;][]{martinez22b}. 
Further observational and theoretical studies are necessary to understand what a normal CCSN is. 

We also investigate the properties of proto-neutron stars left behind the explosion. 
We define the PNS surface at the radius where the mass density is $10^{11}\,{\rm g\,cm}^{-3}$. Their final baryonic masses almost converge in our simulation time and range from $1.37\,\Msun$ (s9) to $2.08\,\Msun$ (s23). 
The number distribution of their gravitational mass weighted by the Salpeter IMF shows a clustering around $\sim 1.4\,\Msun$ and well matches observation except low-mass ($<1.2\,\Msun$) and high-mass ($>2\,\Msun$) ends. These extreme NS masses might originate from binary interaction, which is not considered in this study. 
Note that numerical studies suggest that even the explosion of a very lightweight star, 8--10$\,\Msun$ star with O-Ne-Mg core or ultrastripped star with mass $<2.5\,\Msun$, would not leave behind such a diminutive NS
\citep{kitaura06,suwa15,mueller18}.

Our simulations have started from spherically symmetric initial conditions, except the initially added dipole-like weak magnetic field, and a small density perturbation added at the time of core bounce. 
As hydrodynamics instabilities develops, the distribution of the matter behind the deformed shock becomes highly aspherical and finally ejected materials show a variety of morphological structures. 
This asymmetric explosion leads to a hydrodynamic kick of the protoneutron star. 
The kick velocity derived from our models is smaller (20--250 ${\rm km\,s}^{-1}$ at 0.5 s after bounce) than typical observed values 
\citep[mean values for Galactic single pulsars $\sim 500{\rm km\,s}^{-1}$;][]{kapil23}.
It is still growing at the end of our simulations and larger kick velocities will be obtained by simulations over longer duration. 
In this paper we only consider the NS kick induced by hydrodynamical matter motions. 
Recent 3D CCSN studies suggest that 
recoil due to asymmetric neutrino emission can be a significant factor of the NS kick velocity \citep{coleman22,janka24}. 
The role of the asymmetric neutrino emission of our models in the NS recoil will be investigated in our forthcoming paper.

Spins are also one of the characteristic properties of NSs. 
Our PNS models initiated from non-rotating progenitors acquired spin motions with the period of several hundred milliseconds. 
The central remnants of the s23 and s24 models, showing a sign of spiral SASI mode, obtained the fastest rotating motions with $T_{\rm NS}=40$ -- 60 ms.
They present discontinuous increases in angular momentum even after the shock revival, 
highlighting the importance of sustaining angular momentum injection from the shocked matter to form a rapidly rotating NS 
\citep[see][and references therein for various scenarios of NS spin production]{janka22}.

Though our calculation is based on three-dimensional magnetohydrodynamic simulations with neutrino radiation, much room still remains for quantitative improvement. To handle heavier proto-neutron stars, more careful treatment of general relativity would be necessary \citep{mullerb12a,kuroda22,rahman22,kuroda23}. 
Also, neutrino oscillation may enhance or suppress the neutrino heating \citep{suwa11,ehring23a,ehring23b,nagakura23may}.
Employing such effect would be an urgent task for the prediction of observable.

Recently, effects of binary interaction in the stellar evolution is eagerly investigated \citep{Cantiello2007,Patton2020,Schneider2021,Laplace2021,kinugawa24}.
This is an important update to the previously used single star evolution models \citep{WHW02,woosley07,sukhbold16,Sukhbold18}.
Systematic supernova simulations using binary evolution models should be performed as an extension of this study \citep{vartanyan21,wang24}.

This study focuses on the non-rotating, weakly magnetized progenitor models. Such parameter regions should be enlarged to higher rotation and magnetic fields to consider rapidly rotating or strongly magnetised neutron stars.
In fact, a wide variety of explosion mechanisms appears depending on the parameters of rotation and magnetic fields \citep[e.g.,][]{Iwakami14,summa18,kuroda20,takiwaki21,obergaulinger22,varma23,matsumoto22,bugli23,shibagaki24,hsieh24}.
Our final goal is to depict the fate of massive stars as a landscape of progenitors' dependence. We are taking steady steps toward that goal.

\section*{Acknowledgements}
This study was supported in part by Grants-in-Aid for Scientific Research of the Japan Society for the Promotion of Science (JSPS, Nos. 
JP23K20862, %kiban-b katsuda, nakamura Co-I (-2026.3)
JP23K20848, % kiban-b sotani (-2025.3)
JP23K22494, % kiban-b kotake (-2026.3)
JP23K25895, % Yokoi Kiban B, Takiwaki Co-I (-2028.3)
JP23K03400, % Takiwaki Kiban C (-2026.3)
and JP24K00631) % Yamamoto Kiban B, Takiwaki Kotake Co-I (-2027.3)
the Ministry of Education, Science and Culture of Japan (MEXT, Nos. 
%JP17H06357, %xxx GW genesis soukatsu (Do not cite) xxx
JP17H06364, %GW genesis c01 (-2022.3)
JP17H06365, %GW genesis c02 (-2022.3)
JP19H05811), %nu e02 (-2024.3) 
by the Central Research Institute of Explosive Stellar Phenomena (REISEP) 
and funding from Fukuoka University (Grant Nos. 207002, GR2302),
and JICFuS as “Program for Promoting researches on the Supercomputer Fugaku” (Structure and Evolution of the Universe Unraveled by Fusion of Simulation and AI; Grant Number JPMXP1020230406). 
Numerical computations were in part carried out on Cray XC50 at Center for Computational Astrophysics, National Astronomical Observatory of Japan.

%%%%%%%%%%%%%%%%%%%%%%%%%%%%%%%%%%%%%%%%%%%%%%%%%%
\section*{Data Availability}

The data underlying this article will be shared on reasonable request to the corresponding author.

\bibliographystyle{mnras}
\bibliography{reference,refArXiv}

%\begin{thebibliography}{}
%\expandafter\ifx\csname natexlab\endcsname\relax\def\natexlab#1{#1}\fi
%
%\bibitem[{{Arnaud}(1996)}]{1996ASPC..101...17A}
%{Arnaud}, K.~A. 1996, in Astronomical Society of the Pacific Conference Series,
%  Vol. 101, Astronomical Data Analysis Software and Systems V, ed. G.~H.
%  {Jacoby} \& J.~{Barnes}, 17
%
%
%\end{thebibliography}

%\end{document}

%%%%%%%%%%%%%%%%% APPENDICES %%%%%%%%%%%%%%%%%%%%%

\appendix

\section{Full set of figures}\label{sec:restandfullsetmodels}

In this appendix, we present figures corresponding to those in the main text for all (sixteen) models, including those not depicted in the main body.
For the models which already have been appeared in the main text 
we use the same line style and colors.

%\begin{table}
%\caption{Progenitor properties}
%\begin{tabular}{lccccccc} 
%\hline
%  &$M_{\rm Fe}$ & $R_{\rm Fe}$ &$\xi_{2.0}$ & $\xi_{2.25}$ & $\xi_{2.5}$ &
% $\mu_4$     & $m_4$  \\
%       & [$\Msun$]    & [$10^3$km]  &\multicolumn{4}{c}{[$\Msun / %10^3$km]} &
% [$\Msun$] \\
%\hline
%s9 &  1.32&  1.39&  5e-5 &  4e-5 &  4e-5 &  0.023&  1.36\\
%s10&  1.37&  1.31&  0.007&  0.004&  2e-4 &  0.073&  1.42\\
%s11&  1.37&  1.32&  0.022&  0.012&  0.008&  0.056&  1.41\\
%s12&  1.44&  1.77&  0.092&  0.037&  0.022&  0.078&  1.44\\
%s13&  1.50&  1.65&  0.234&  0.131&  0.058&  0.117&  1.61\\
%s14&  1.51&  1.81&  0.254&  0.172&  0.123&  0.113&  1.69\\
%s15&  1.49&  1.91&  0.280&  0.210&  0.166&  0.153&  1.44\\
%s16&  1.45&  1.58&  0.232&  0.183&  0.154&  0.090&  1.51\\
%s17&  1.46&  1.53&  0.244&  0.193&  0.164&  0.100&  1.53\\
%s18&  1.50&  1.77&  0.252&  0.200&  0.171&  0.110&  1.50\\
%s19&  1.54&  1.73&  0.311&  0.222&  0.177&  0.103&  1.64\\
%s20&  1.61&  1.81&  0.422&  0.312&  0.260&  0.118&  1.80\\
%s21&  1.44&  1.55&  0.202&  0.159&  0.138&  0.098&  1.49\\
%s22&  1.63&  1.85&  0.459&  0.335&  0.274&  0.118&  1.81\\
%s23&  1.83&  2.24&  0.714&  0.536&  0.426&  0.179&  2.12\\
%s24&  1.81&  2.21&  0.697&  0.509&  0.408&  0.173&  2.08\\
%s25&  1.71&  2.01&  0.564&  0.404&  0.337&  0.159&  1.92\\
%\hline
%\multicolumn{8}{l}{All values are estimated from the pre-collapse progenitor data.}\\
%\label{tbl:prog}
%\end{tabular}
%\end{table}
% from MBP:Documents/0_research/labo/databox/progenitor/prog_sukhbold16/mktbl

\begin{figure}
    \includegraphics[width=0.95\linewidth]{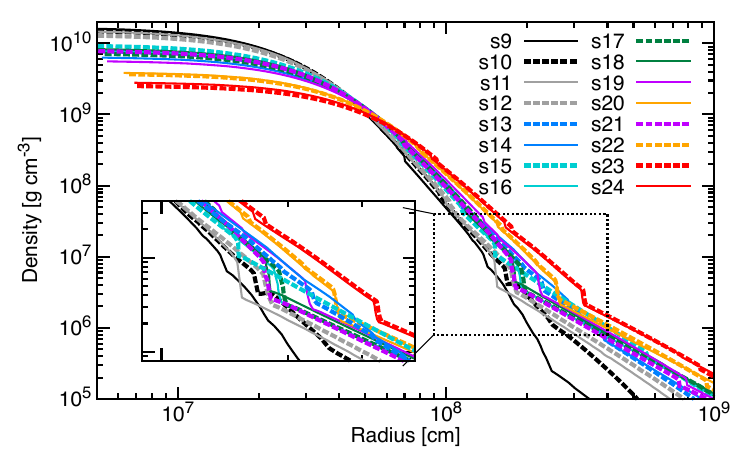} 
    \caption{
    Density distribution of the sixteen progenitor models examined in this paper. The inset panel highlights the profile around the density drops located at 1500--3000 km, corresponding to the interface between Fe and Si/O layers.
    }
    \label{fig:prog-den-all}
\end{figure}

\begin{figure}
    \begin{center}
    \includegraphics[width=0.95\linewidth]{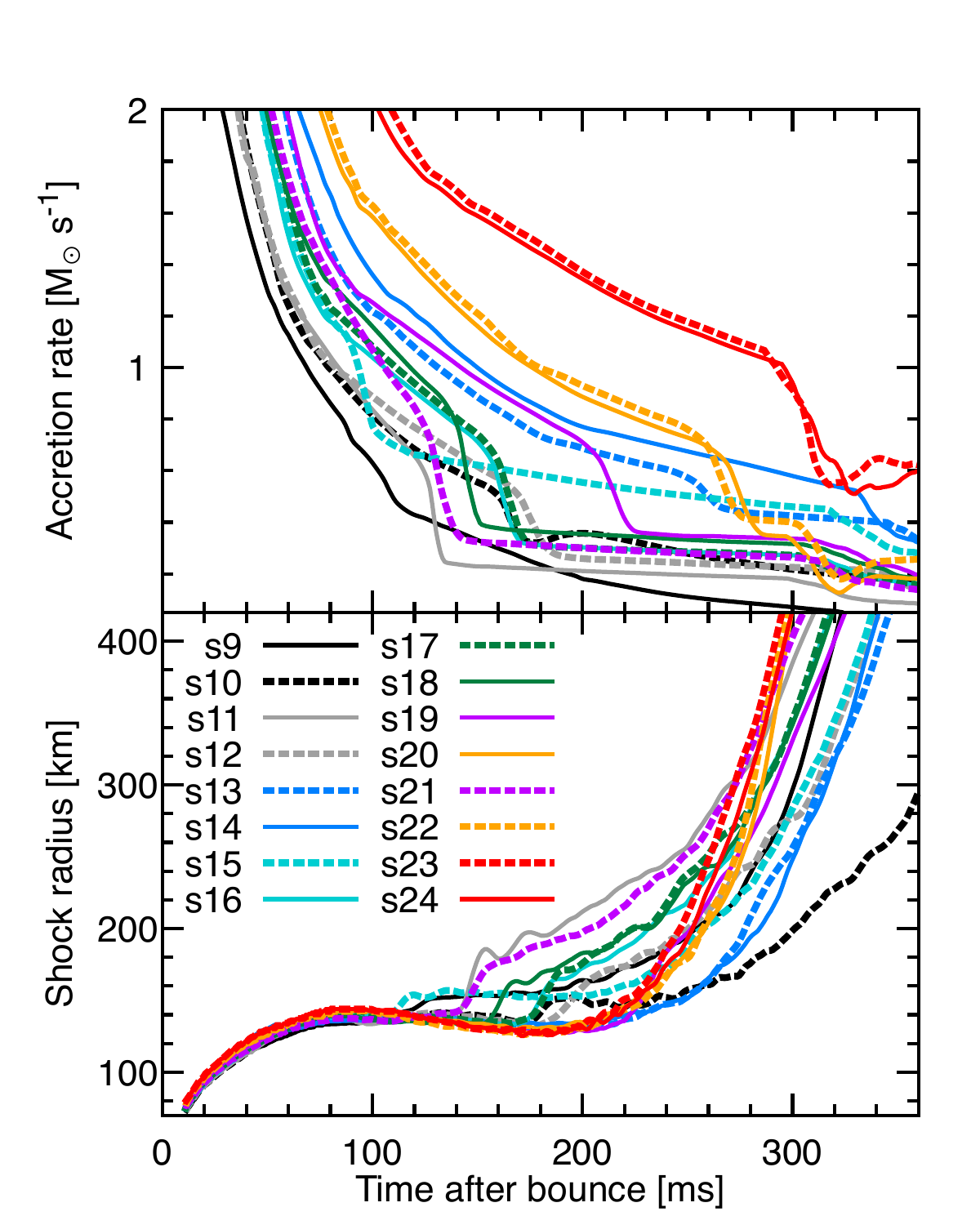}
    \end{center}
    \caption{
    Time evolution of the mass accretion rate at 500 km in radius (top panel) and the angle-averaged shock radius (bottom panel) in the early 360 ms. 
    Most of the models show a drop in the accretion rate and a corresponding jump in the shock radius when the Si/O interface falls onto the central region (Figure \ref{fig:prog-den}). 
    The s14 model (blue solid line) does not hold a sharp density decline at the Si/O interface. 
    The s13 model (blue dashed line) exhibits a reduction in the accretion rate at $\sim 250$ ms after bounce, but the magnitude of this reduction is comparatively smaller when contrasted with other models.
    These models take a relatively longer time for the shocks to turn to the expansion.
    Note that the drops of the accretion rate later than 300\,ms after bounce are caused by the expanding shock waves.
    }
    \label{fig:mdot-rsh-all}
\end{figure}

Figure \ref{fig:prog-den-all} is the complete version of Figure \ref{fig:prog-den}, showing the density profiles of the sixteen progenitor models examined in this paper. 
Most of the progenitors show a clear density drop at 1500--3000 km, except the models s13 and s14. 
The peculiar density structure of these two models are reflected to their higher accretion rates and later shock revival time compared with the other models starting from more massive progenitors (Figure \ref{fig:mdot-rsh-all}).

\begin{figure}
    \begin{center}
    \includegraphics[width=0.98\linewidth]{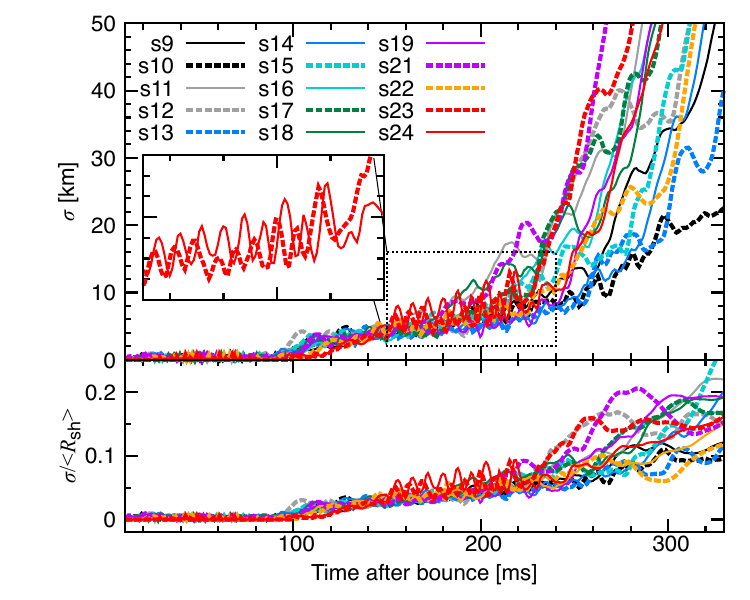}
    \end{center}
    \caption{Rms deviation $\sigma$ of the shock radius from its angle-averaged value. The inset in the top panel extracts the $\sigma$ of the s23 and s24 models in 150--240 ms post bounce. They present a quasi-periodic oscillation in this phase, corresponding to the spiral SASI motion of the shock front. The ratio of the $\sigma$ to the angle-averaged shock radius (bottom panel) is less than 10 percent before shock revival.}
    \label{fig:sigma-all}
\end{figure}

Figure \ref{fig:sigma-all} is a complete version of Figure \ref{fig:sigma}, showing rms deviation $\sigma$ of the shock radius from its angle-averaged value. 
The most notable feature is a quasi-periodic oscillation presented by the models s23 and s24 (red lines, emphasized in the inset), which can be attributed to the spiral mode of SASI motions. 
All models, including s23 and s24, exhibit small $\sigma$ ($<10$\%) before the shock revival and we conclude that our 3D models do not show any strong sloshing mode of SASI motions. 

\begin{figure*}
    \begin{center}
{
    \tabcolsep = 1pt
 %   \tabrowsep = 1pt
    \begin{tabular}{cccc}
    \includegraphics[width=0.23\linewidth]{./fig-snap/s9-En3wunst-00635.jpeg} &
    \includegraphics[width=0.23\linewidth]{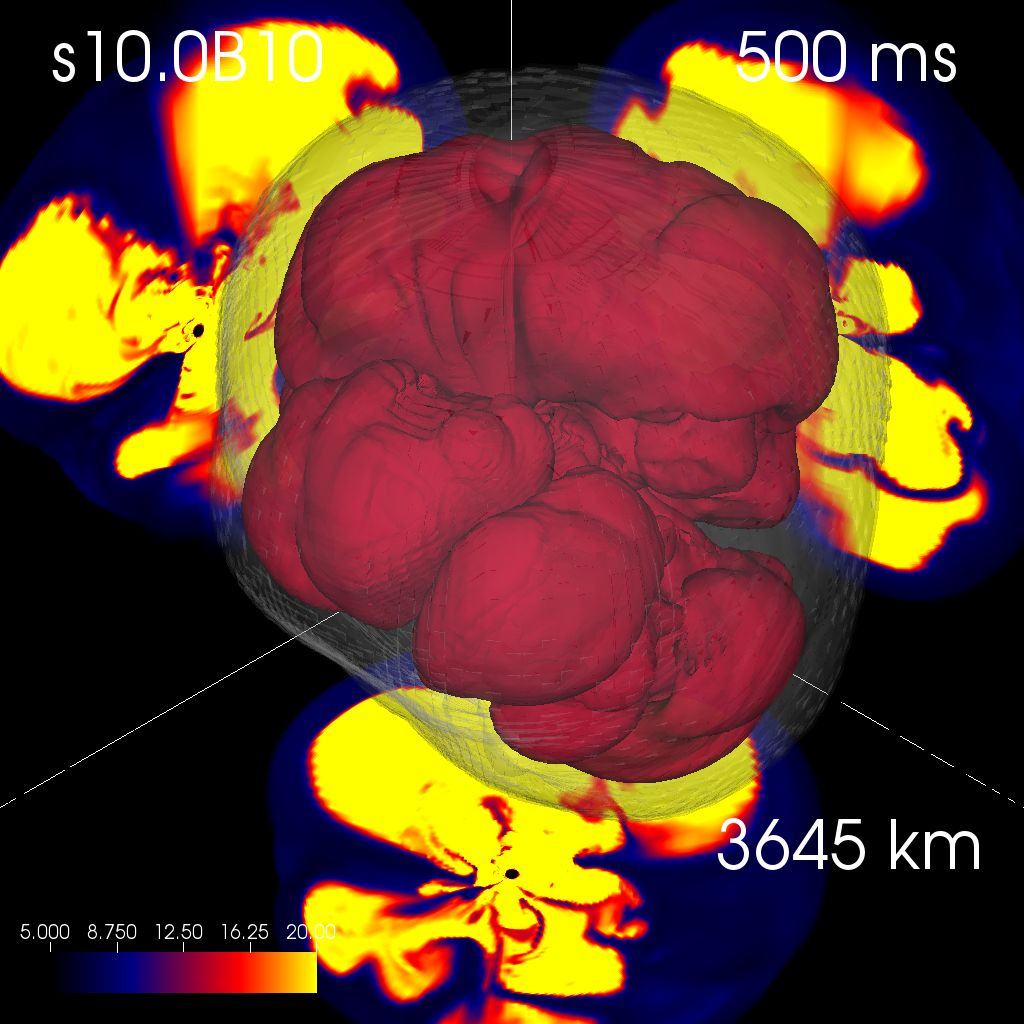} &
    \includegraphics[width=0.23\linewidth]{./fig-snap/s11-En3wunst-00623.jpeg} &
    \includegraphics[width=0.23\linewidth]{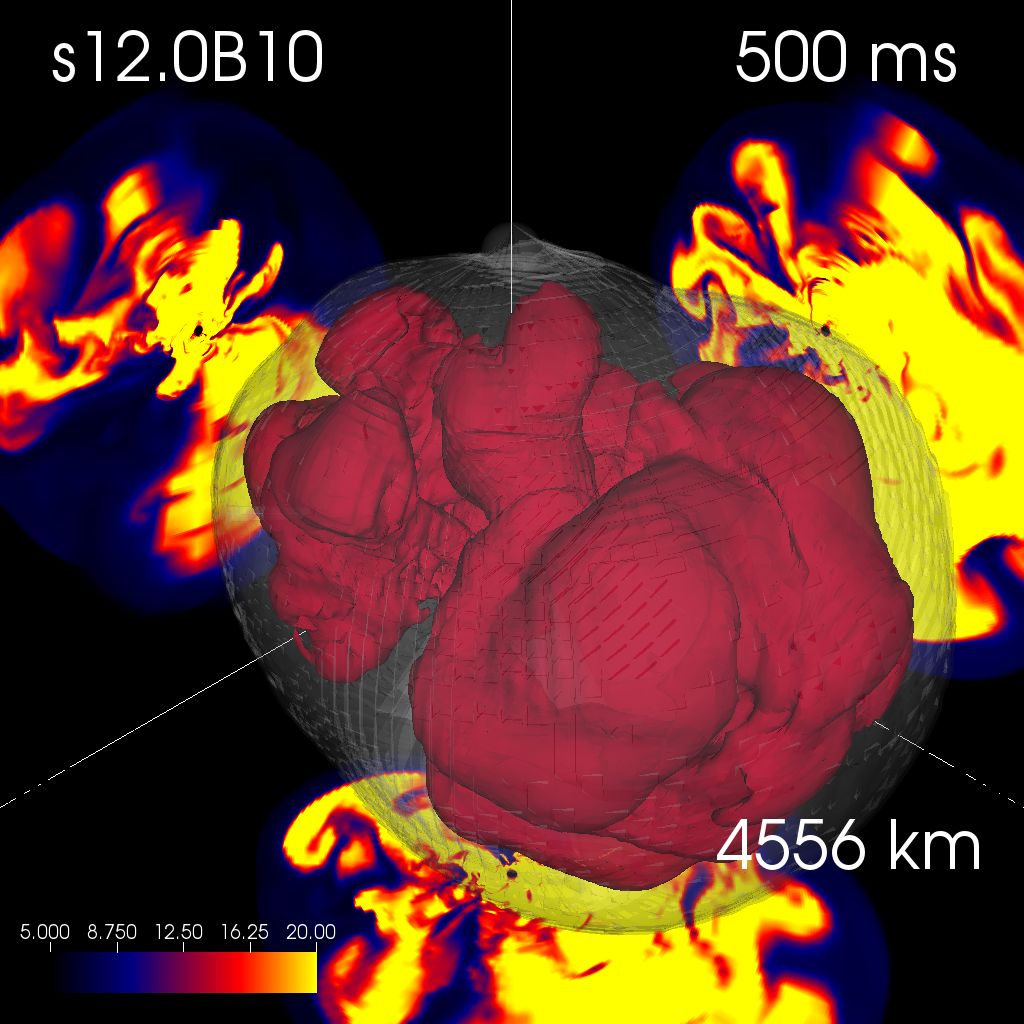} \\
    \includegraphics[width=0.23\linewidth]{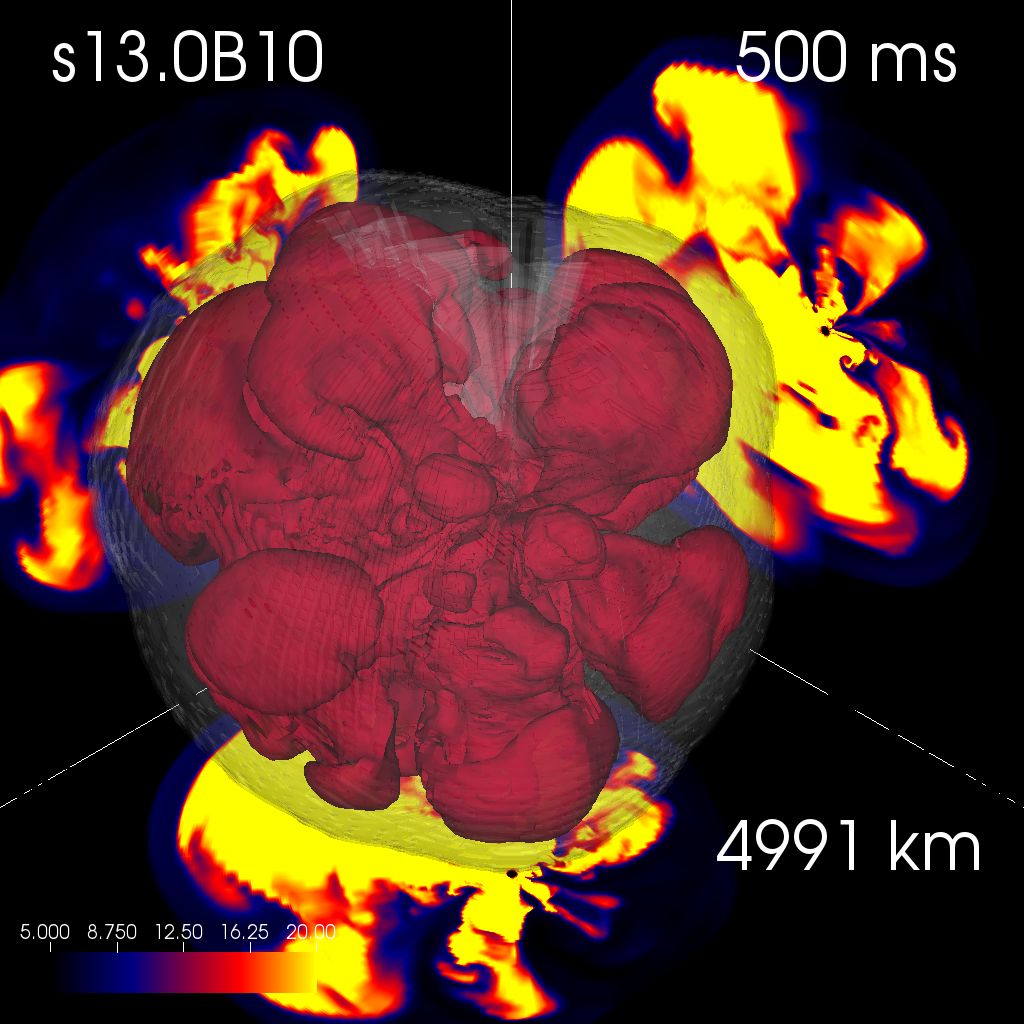} &
    \includegraphics[width=0.23\linewidth]{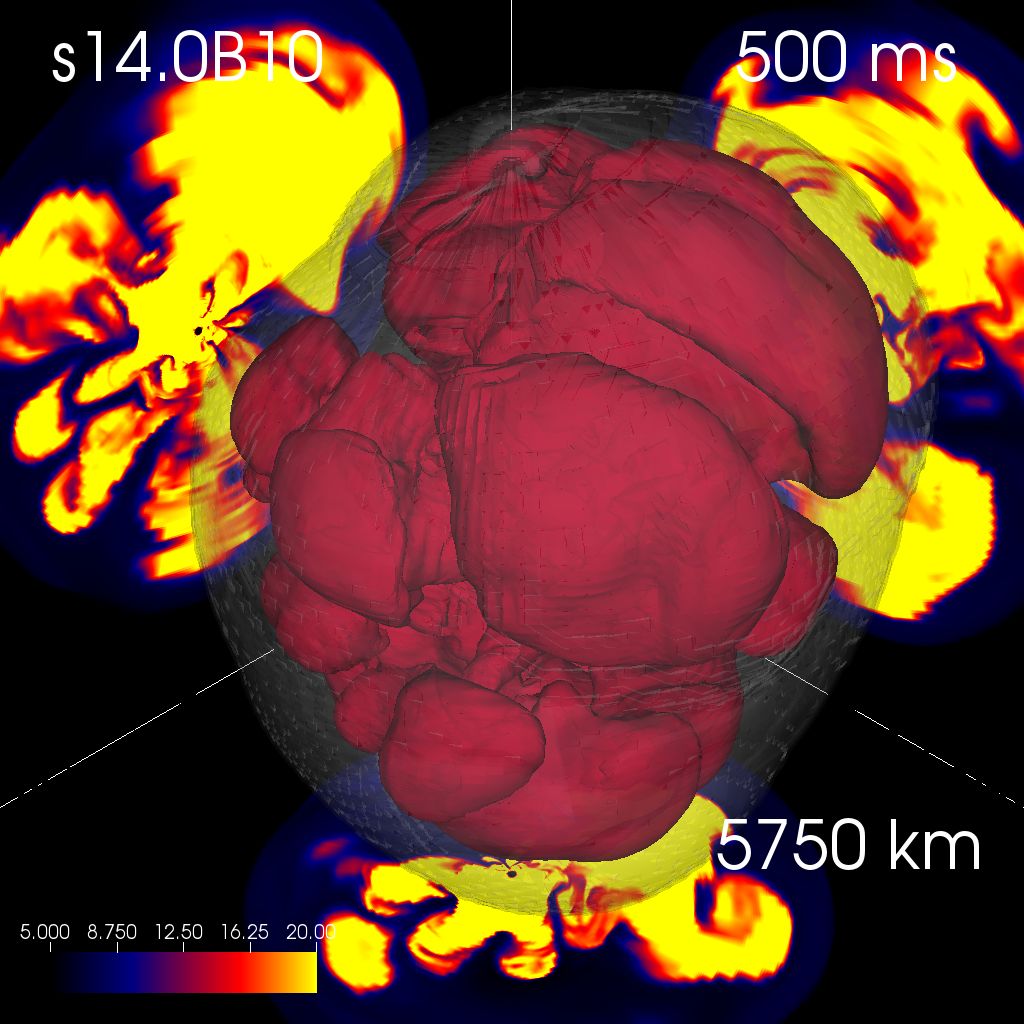} &
    \includegraphics[width=0.23\linewidth]{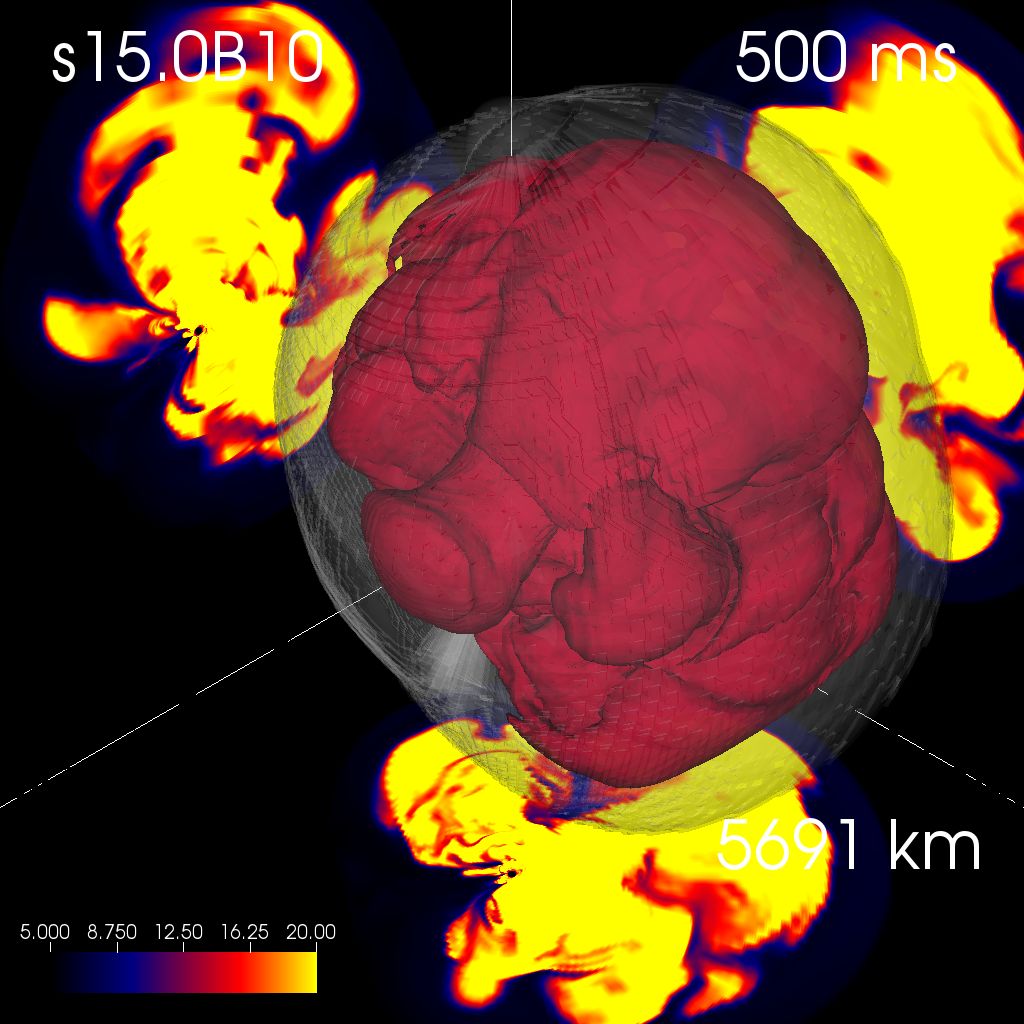} &
    \includegraphics[width=0.23\linewidth]{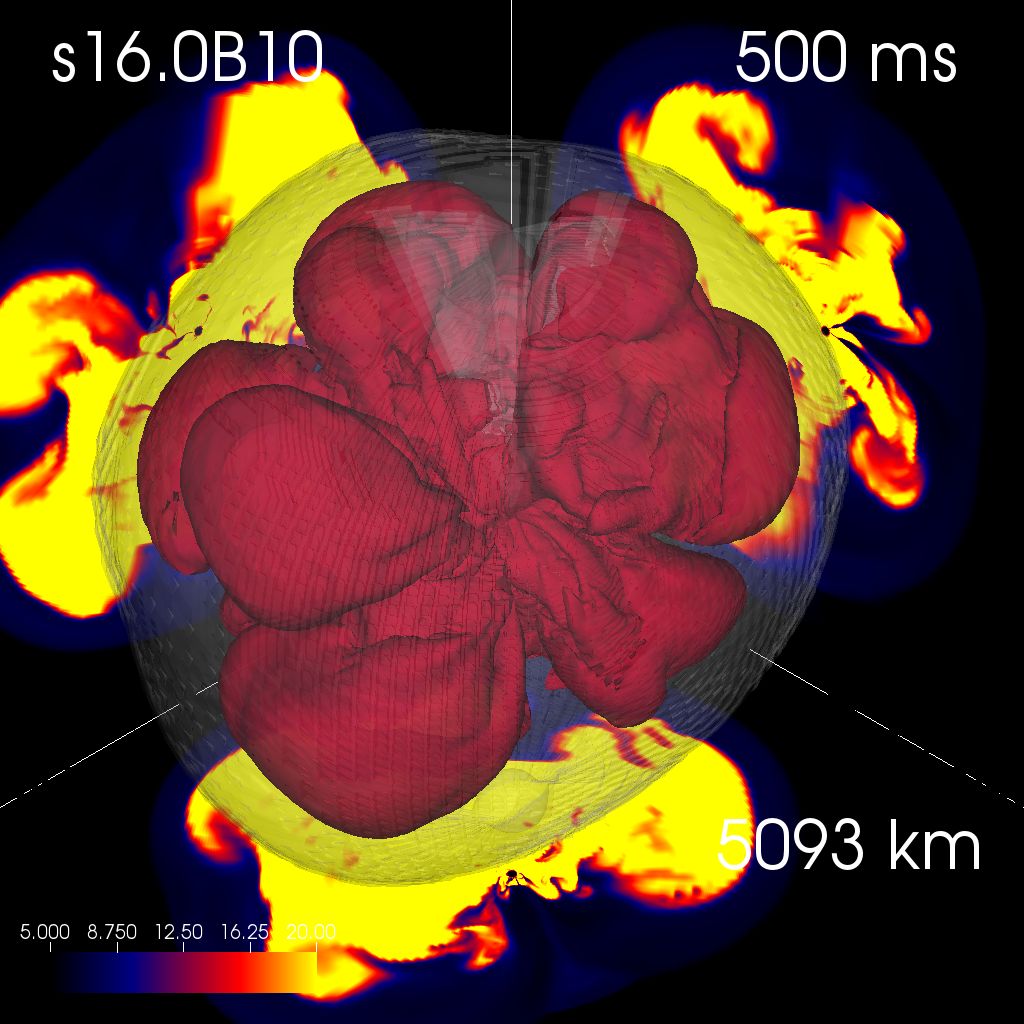} \\
    \includegraphics[width=0.23\linewidth]{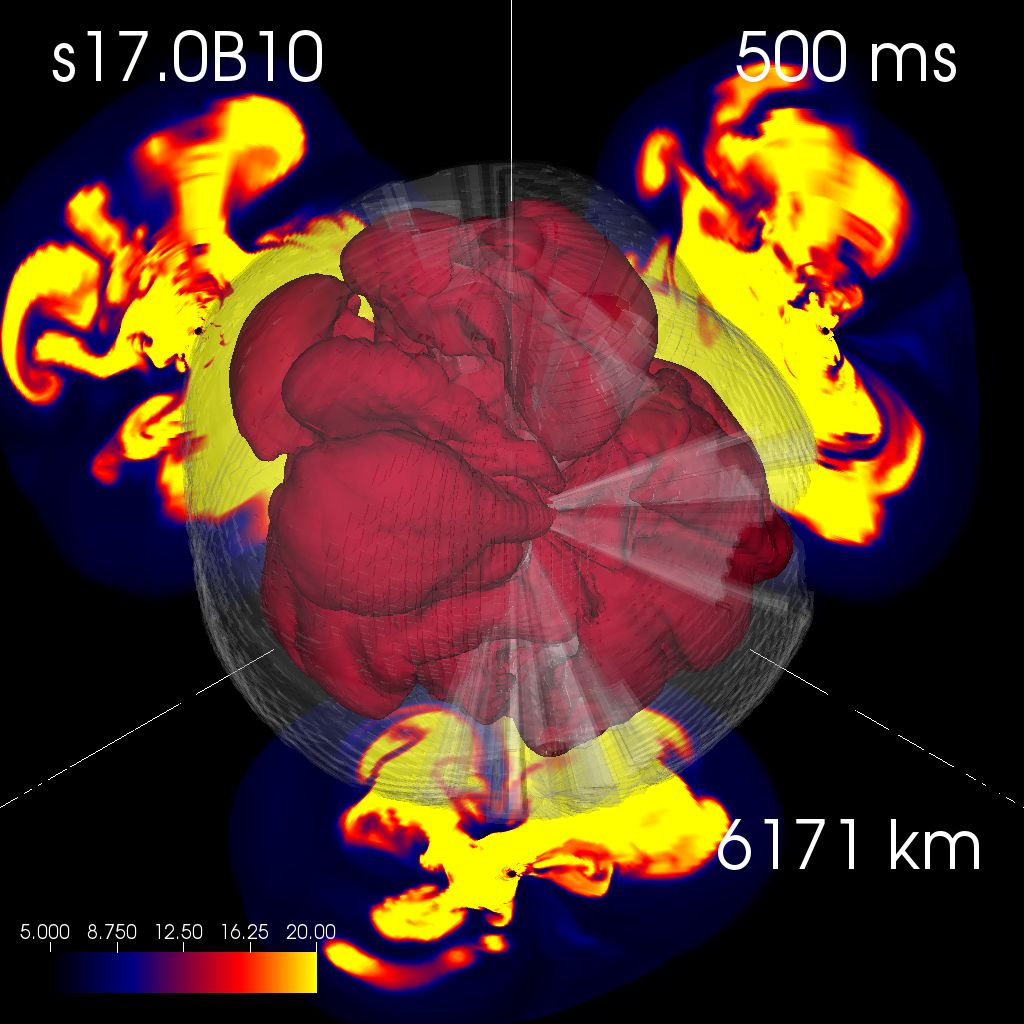} &
    \includegraphics[width=0.23\linewidth]{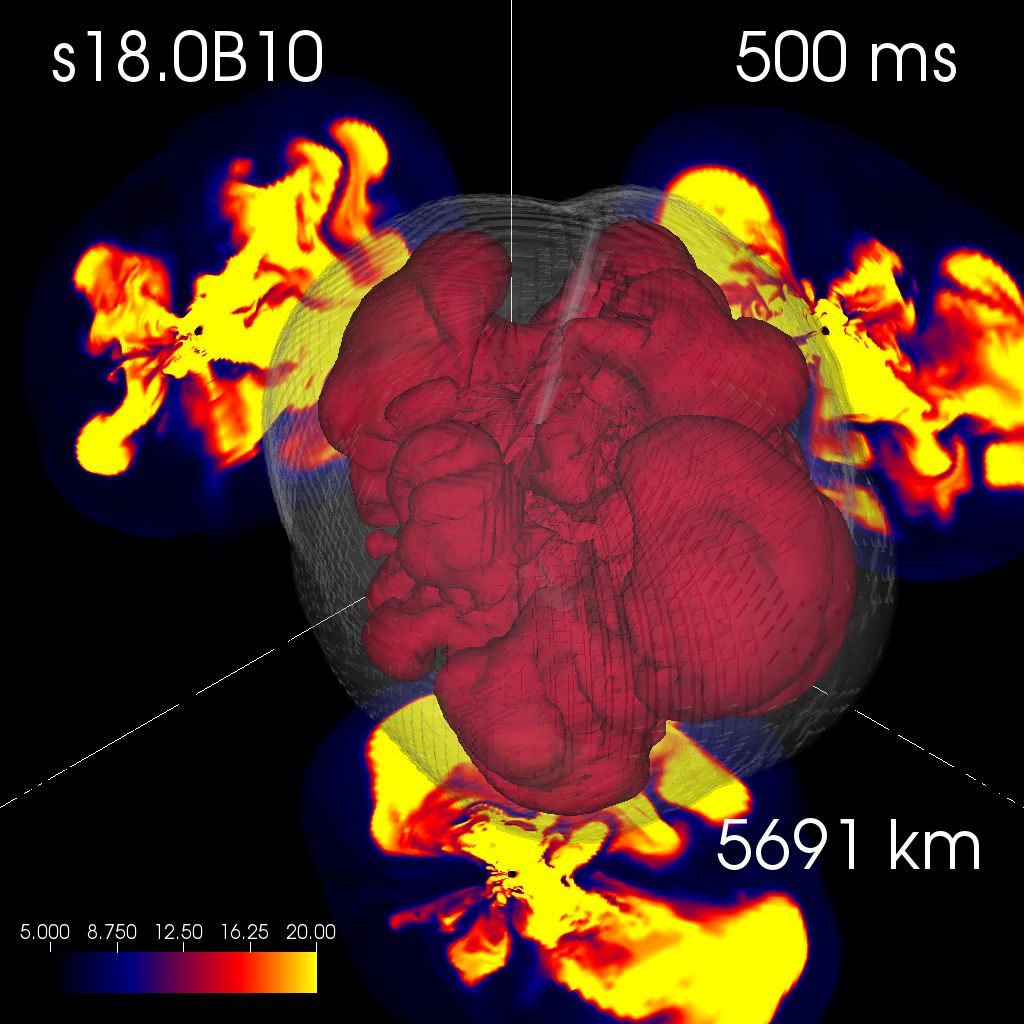} &
    \includegraphics[width=0.23\linewidth]{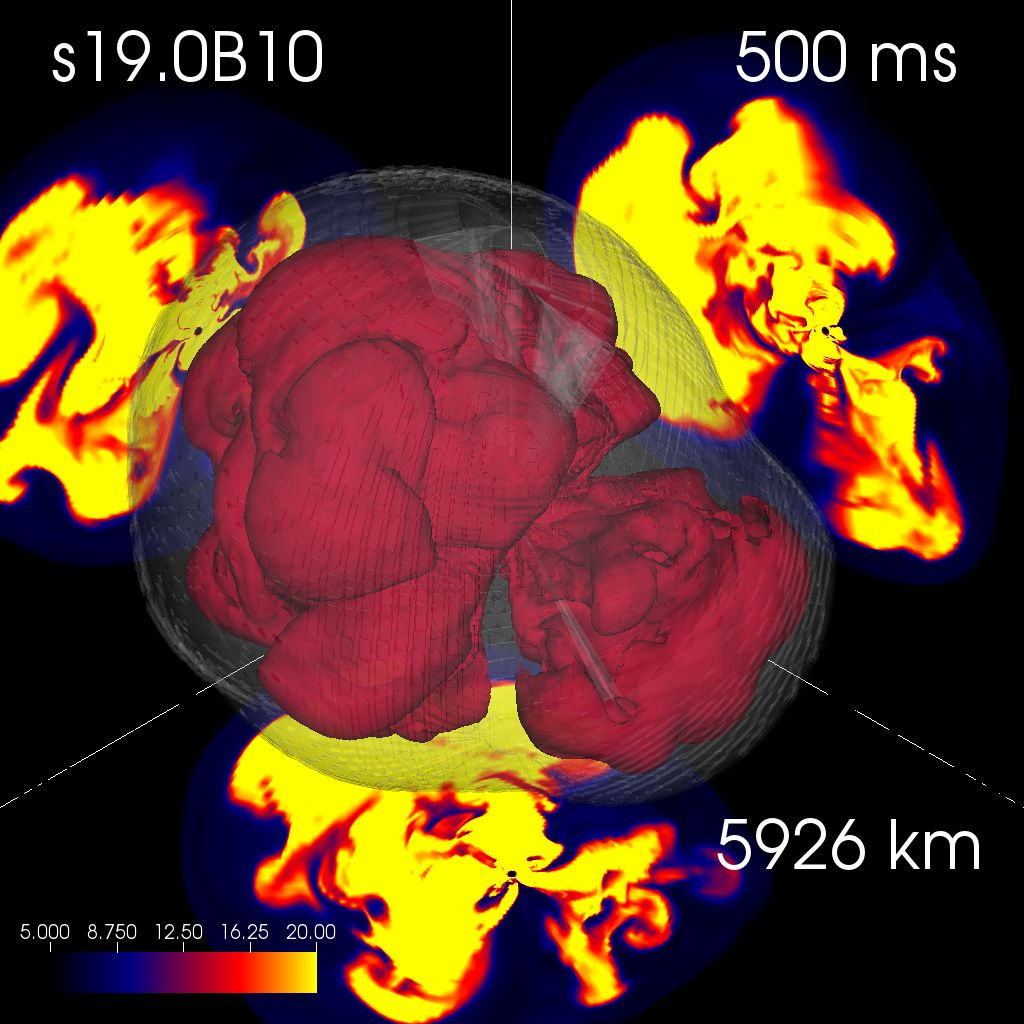} &
    \includegraphics[width=0.23\linewidth]{./fig-snap/s20-En3wunst-00733.jpeg} \\
    \includegraphics[width=0.23\linewidth]{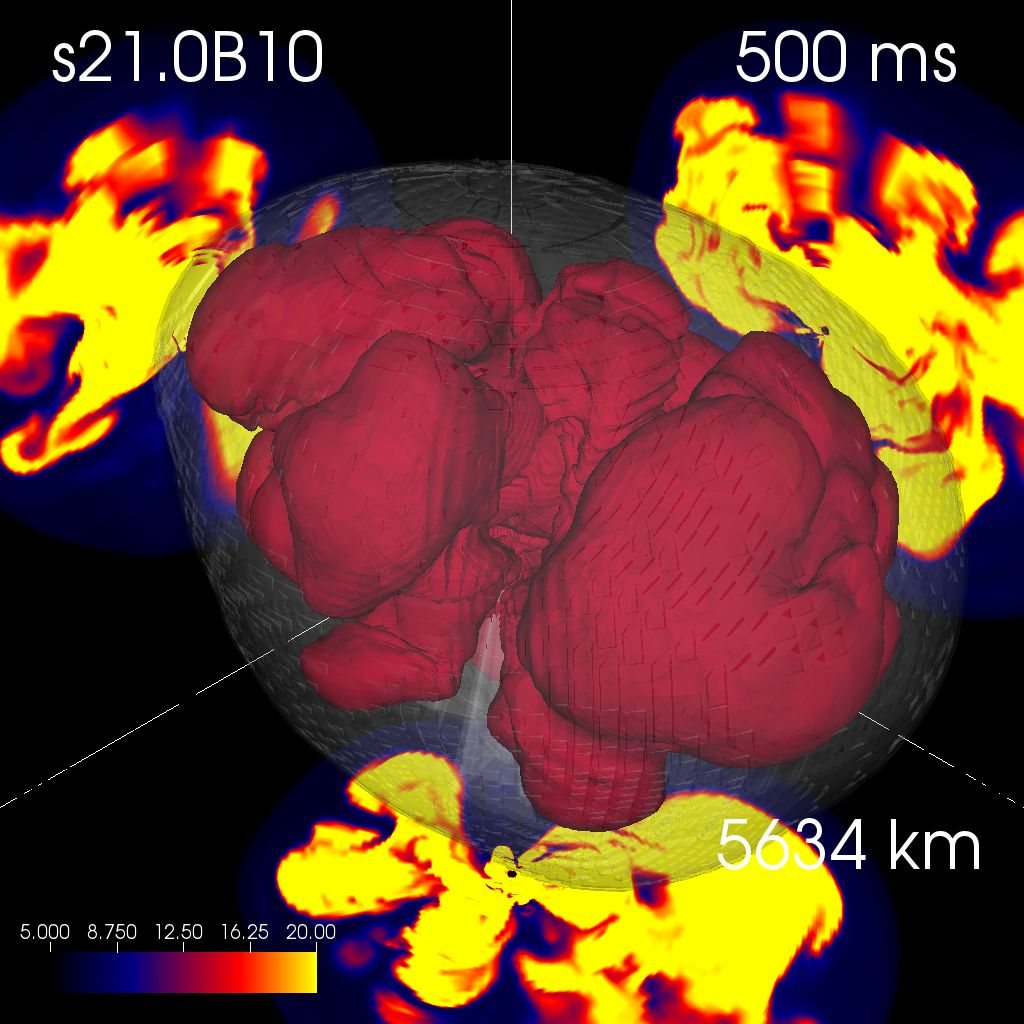} &
    \includegraphics[width=0.23\linewidth]{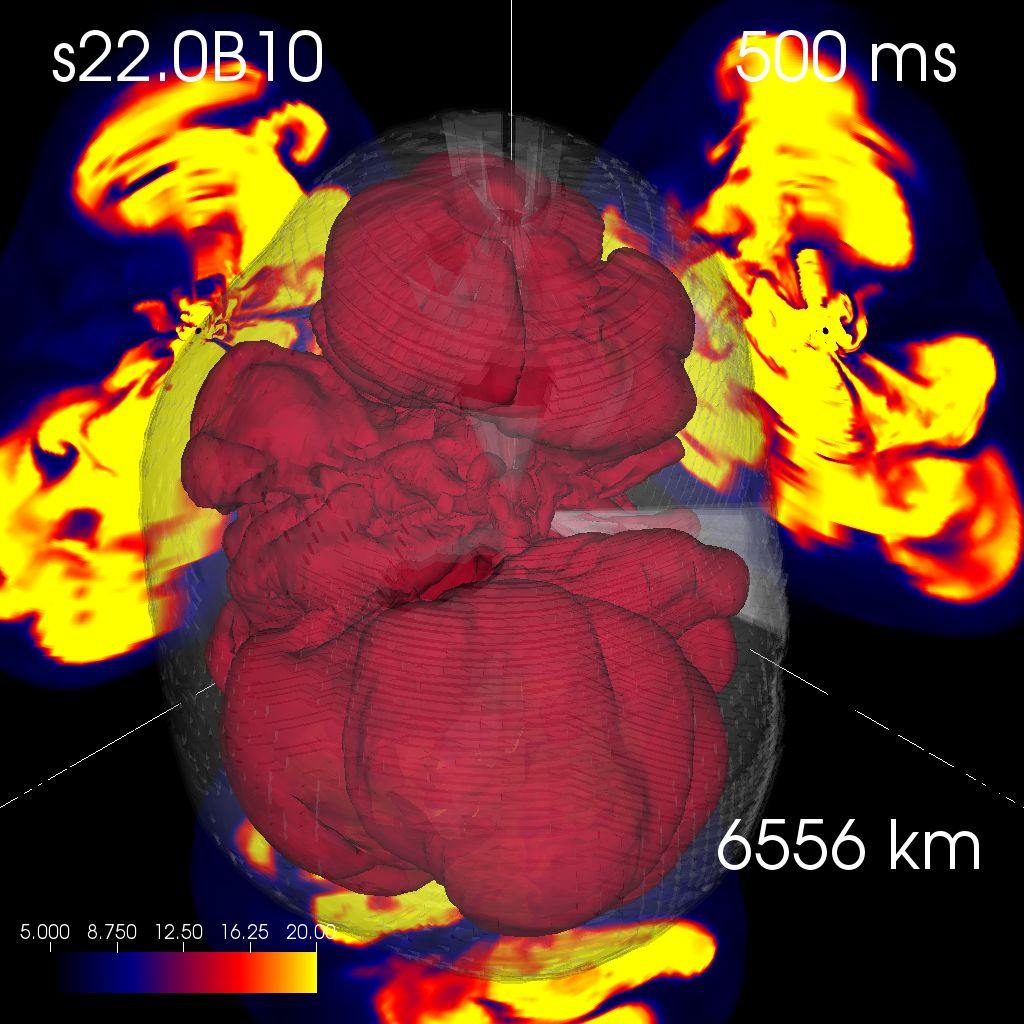} &
    \includegraphics[width=0.23\linewidth]{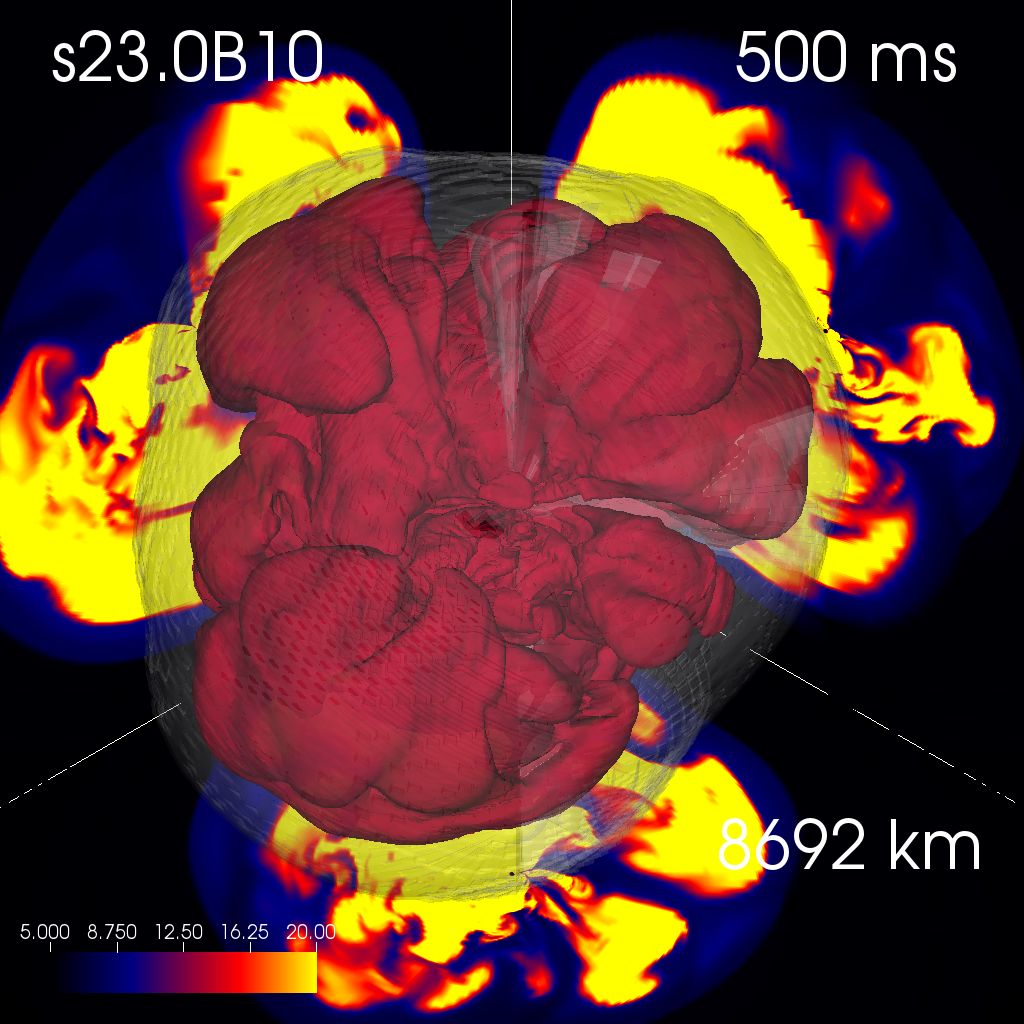} &
    \includegraphics[width=0.23\linewidth]{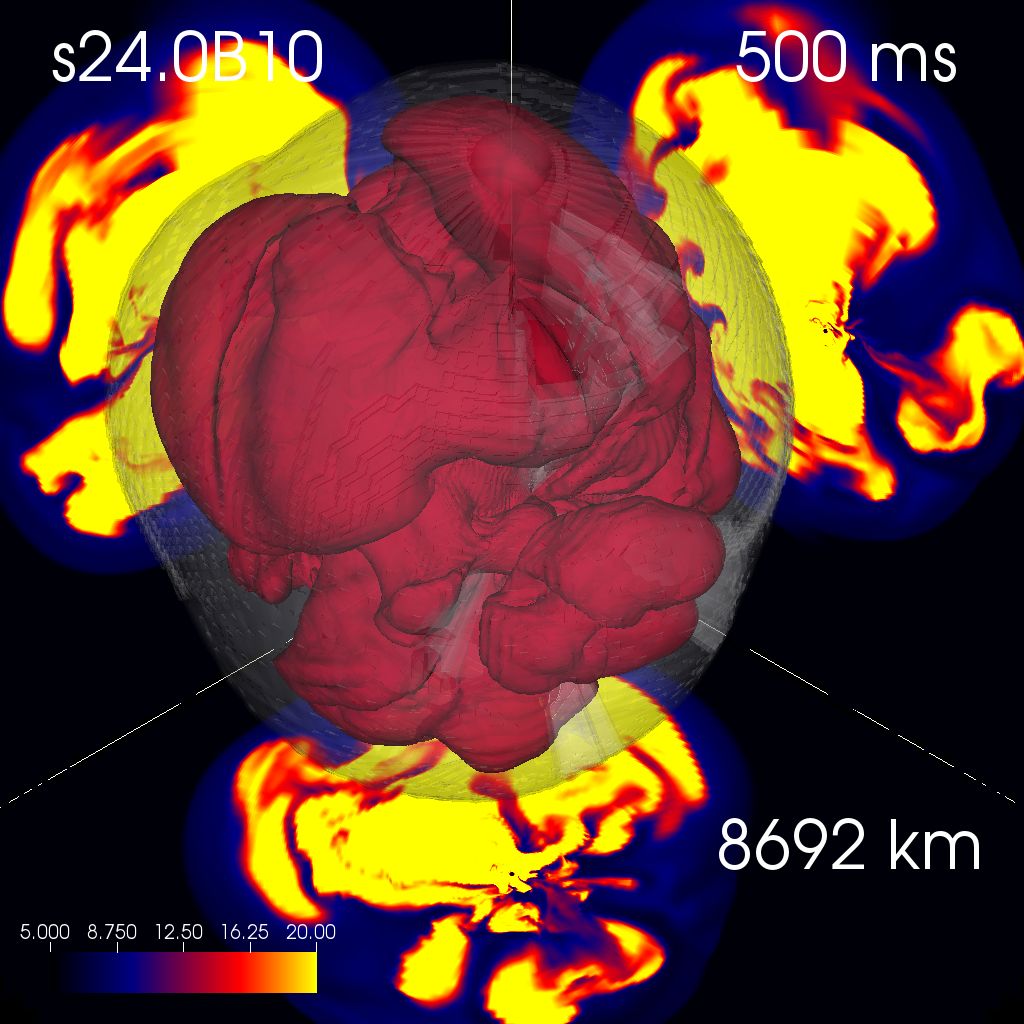} \\
    \end{tabular}
    }
    \end{center}
    \caption{Snapshots of the isosurface of entropy at 500 ms after bounce for the sixteen models from s9 (top-left panel) to s24 (bottom-left panel) in order of the progenitor's ZAMS mass. The scale of the visualized box is shown in the bottom-right corner in each panel. The shock radius at the end of our simulations is more than several thousands of kilometres and we find that all models examined in this study present successful shock revival. The contours on the cross-sections in the $x=0$, $y=0$, and $z=0$ planes are projected on the walls at the back right, back left, and bottom of each panel, respectively.
    The models s19 and s20 show a characteristic bipolar-like matter distribution with a ``neck'' around the central proto-neutron star where low-entropy down streams exits.}
    \label{fig:snap-all}
\end{figure*}

Figure \ref{fig:snap-all} shows snapshots of the isosurface of entropy at 500 ms after bounce for the models examined in this paper. 
The panels of the models s9, s11, and s20 are the same as those shown in the main text (Figure \ref{fig:snap}). 
They have some common features: 
expanding shock surfaces located at several thousand kilometers at 0.5 s after bounce, large scale high-entropy blobs behind the shock waves, and low-entropy downflows between them. 
The models s19 and s20 show a characteristic bipolar-like matter
distribution. This features could be related to double-peaked line emission in the nebular spectroscopy \citep{Fang24} and earlier emergence of the polarization in the supernova observations \citep{Nagao24}.

\begin{figure*}
    \begin{center}
    \begin{tabular}{cc}
    \includegraphics[width=0.45\linewidth]{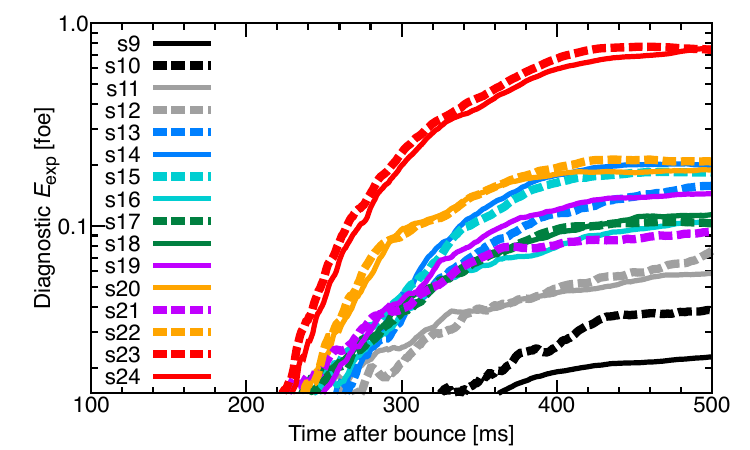} &
    \includegraphics[width=0.45\linewidth]{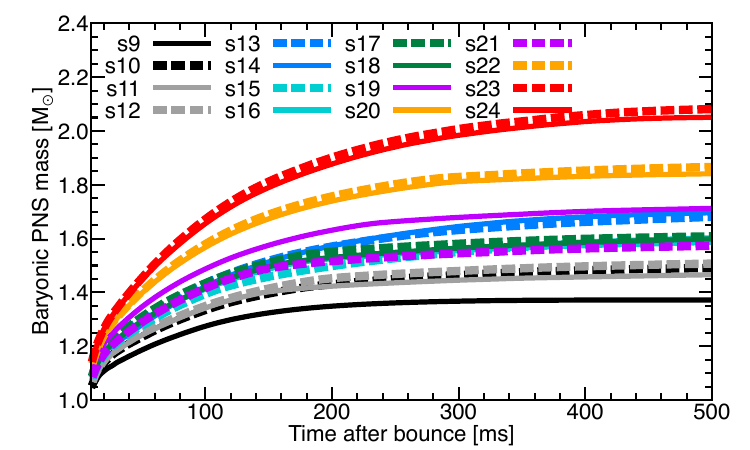} \\
    \includegraphics[width=0.45\linewidth]{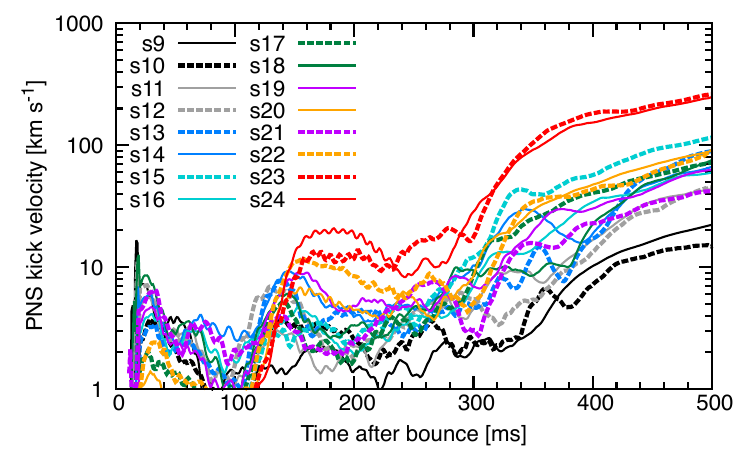} &
    \includegraphics[width=0.45\linewidth]{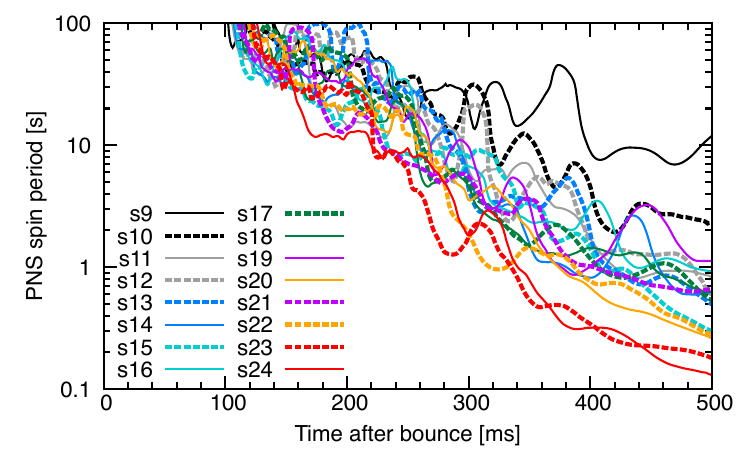}
    \end{tabular}
    \end{center}
    \caption{Time evolution of the explosion energy and PNS properties. 
    Top-left: time evolution of the diagnostic explosion energy. Most models present 0.1--0.2 foe at the final simulation time, except some small mass models (s9 and s10, $E_{\rm exp}<0.04$ foe) and large mass models with energetic explosion (s23 and s24, $E_{\rm exp}\sim$ 0.7--0.8 foe).
    Top-right: time evolution of the baryonic PNS mass.
    Bottom-left: Time evolution of the PNS kick velocity. At the final time of our simulations, the kick velocity is mainly in the range $\sim$ 20--100 km s$^{-1}$, whereas the fastest ones (the models s23 and s24 shown by red lines) exhibit $\sim 250$ km s$^{-1}$.
    Bottom-right: Time evolution of the PNS spin period. 
    To improve readability, the lines in the bottom panels have each been smoothed using a 20-ms moving average.}
    \label{fig:eexp-all}
\end{figure*}

Figure \ref{fig:eexp-all} summarizes the explosion energy and PNS properties of our 3D magnetic CCSN models. 
As a general trend, massive progenitors (s23 and s24, red lines) show energetic explosions, leaving rapidly-rotating massive PNSs with large kick velocities, and small mass progenitors (s9 and s10, black lines) are in opposite. 

%\begin{figure}
%    \includegraphics[width=0.95\linewidth]{./fig-pns/mpns.pdf} 
%    \caption{Mass properties of the central remnants. 
%    Left panel: time evolution of the baryonic PNS mass for the sixteen %models with magnetic fields.}
%    \label{fig:pns-mass-all}
%\end{figure}

%\begin{figure}
%    \begin{center}
%    \includegraphics[width=0.95\linewidth]{./fig-pns/ns_kick_tav.pdf}
%    \end{center}
%    \caption{Time evolution of the PNS kick velocity. At the final time of our simulations, the kick velocity is mainly in the range $\sim$ 20--100 km s$^{-1}$, whereas the fastest ones (the models s23 and s24 shown by red lines) exhibit $\sim 250$ km s$^{-1}$.
%    To improve readability, the lines have each been smoothed using a 20-ms moving average.}
%    \label{fig:kick-all}
%\end{figure}

%\begin{figure}
%    \begin{center}
%    \includegraphics[width=0.95\linewidth]{./fig-pns/ns_spin_tav.pdf}
%    \end{center}
%    \caption{Time evolution of the PNS spin period. 
%    To improve readability, the lines have each been smoothed using a 20-ms moving average.}
%    \label{fig:spin-all}
%\end{figure}

%\begin{figure}
%    \includegraphics[width=0.95\linewidth]{./fig-bfld/ns_b.pdf}
%    \caption{Time evolution of the total magnetic strength at the PNS surface.
%    }
%    \label{fig:r-b-all}
%\end{figure}

% Don't change these lines
\bsp	% typesetting comment
\label{lastpage}
\end{document}